\documentclass[twocolumn,aps,showpacs,amssymb,floatfix]{revtex4}
\newcommand{\be}{\begin{equation}}  
\newcommand{\ee}{\end{equation}}  
\newcommand{\beq}{\begin{eqnarray}}  
\newcommand{\eeq}{\end{eqnarray}}  
\usepackage{bm}% bold math  
\usepackage{graphicx}%  
\usepackage{epsf}  
\usepackage{epsfig}  
   
 \usepackage{amsmath}  
\usepackage{amsfonts,amssymb}

\usepackage{dsfont}  
\usepackage{slashed}  
\usepackage{booktabs}

\begin{document}  
\preprint{DESY 10-216}

   \title{Nucleon electromagnetic form factors in twisted mass lattice QCD}

\author{C.~Alexandrou~$^{(a,b)}$, M. Brinet~$^{(c)}$, J.~Carbonell~$^{(c)}$ M. Constantinou~$^{(a)}$,  P.~A.~Harraud~$^{(c)}$,  
P.~Guichon~$^{(d)}$,   
K.~Jansen~$^{e}$, T.~Korzec~$^{(a,f)}$, M.~Papinutto~$^{(c)}$ }  
\affiliation{$^{(a)}$ Department of Physics, University of Cyprus, P.O. Box 20537, 1678 Nicosia, Cyprus\\  
 $^{(b)}$ Computation-based Science and Technology Research  
    Center, Cyprus Institute, 20 Kavafi Str., Nicosia 2121, Cyprus \\  
$^{(c)}$ Laboratoire de Physique Subatomique et Cosmologie,  
               UJF/CNRS/IN2P3, 53 avenue des Martyrs, 38026 Grenoble, France\\  
$^{(d)}$ CEA-Saclay, IRFU/Service de Physique Nucl\'eaire, 91191 Gif-sur-Yvette, France\\  
$^{(e)}$ NIC, DESY, Platanenallee 6, D-15738 Zeuthen, Germany\\  
         \vspace{0.2cm}  
 $^{(f)}$ Institut f\"ur Physik  
   Humboldt Universit\"at zu Berlin, Newtonstrasse 15, 12489 Berlin, Germany}

 \begin{abstract}

We present results on the  nucleon electromagnetic  form factors within  
lattice QCD using two flavors of degenerate twisted mass  
fermions. Volume effects are   
 examined using simulations at two volumes of spatial length $L=2.1$~fm  
and $L=2.8$~fm.   
Cut-off effects  are investigated using three different  
values of the lattice spacings, namely $a=0.089$~fm, $a=0.070$~fm and $a=0.056$~fm.  
The nucleon magnetic moment, Dirac and Pauli radii  are obtained in the  
continuum limit and chirally extrapolated to  
the physical pion mass  
%%%TOM: enabling -> allowing for a  
allowing for a comparison with experiment.

 \end{abstract}  

\begin{minipage}{0.5\linewidth}  
\end{minipage}\hfill

% \begin{figure}[h]  
%   \begin{center}  
%      \includegraphics[width=0.33\linewidth]{ETMC_Logo.ps}  
%    \end{center}  
%  \end{figure}  

\maketitle

\setcounter{figure}{\arabic{figure}}  
  
\newcommand{\twopt}[5]{\langle G_{#1}^{#2}(#3;\mathbf{#4};\Gamma_{#5})\rangle}  
\newcommand{\threept}[7]{\langle G_{#1}^{#2}(#3,#4;\mathbf{#5},\mathbf{#6};\Gamm  
a_{#7})\rangle}  
  
\newcommand{\Op}{\mathcal{O}} % Fractur O  
\newcommand{\C}{\mathcal{C}} % Fractur C  
\newcommand{\eins}{\mathds{1}} % Fractur O  
  
\bibliographystyle{apsrev}                     % Style for bibliogrpahy  

\section{Introduction}  
Understanding the structure of the nucleon using the  
%%%TOM: underline -> underlying  
underlying theory  
of the strong interactions is a central problem of hadronic physics.  
 The nucleon (N) electromagnetic  
 form factors  provide an indispensable probe to the structure of the   
nucleon.  
 Experiments to measure the electromagnetic nucleon form factors  
have been carried out since the 50's.  
A new generation  
of experiments using polarized beams  
revealed unexpected results~\cite{Jones:1999rz,Gayou:2001qd}.   
The form factors obtained in these  
polarization experiments   
%%%TOM: differed -> differ  
differ from those extracted in previous  
experiments  
%%%TOM: using -> based on   
based on the Rosenbluth cross-section separation method.  
The new  
%%%TOM: inserted 'generation of', showed -> has shown  
generation of experiments has shown that the ratio of the proton electric to  
magnetic form factor $G_E^p/G_M^p$ decreases almost linearly with  
%%%TOM: the momentum -> momentum  
increasing  momentum  
transfer squared instead of being approximately constant.  
Two-photon exchange effects, previously neglected, were shown to be  
the source of the discrepancy. For a recent review we  
refer the reader to Ref.~\cite{Perdrisat:2006hj,Guichon:2003qm}.  
%%%TOM: is it really established that the whole effect is due to 2 photon exchange?  
Precision experiments are currently under way at major facilities in order  
to measure the nucleon form factors   
%%%TOM: 'even' inserted  
even more accurately  
and  at higher values of the momentum transfer~\cite{deJager:2008zza}.  
  
In this work we present results on electromagnetic form factors obtained using two degenerate  
 light quarks  ($N_F{=}2$) in the twisted mass formulation.  
Twisted mass fermions (TMF)~\cite{Frezzotti:2000nk,Shindler:2007vp}  provide an attractive  formulation of lattice QCD that  
allows automatic ${\cal O}(a)$ improvement, infrared regularization   
of small  
eigenvalues and fast dynamical   
simulations~\cite{Frezzotti:2003ni}. For the  
calculation of the nucleon form factors, which is the aim of this work,  
the automatic    
 ${\cal O}(a)$ improvement is particularly relevant   
since it is achieved  by tuning only one parameter in the action,  
requiring no further improvements on the operator level.

The action for two degenerate flavors of quarks  
 in twisted mass QCD is in lattice units given by  
%%%TOM: inserted 'in lattice units' and removed a^4. That's because  
%%%     we later use lattice units in the remaining paper e.g in the  
%%%     definition of the correlators.  
   \begin{equation}  
      S=S_g + \sum_x \bar\chi(x) \left[D_W  
 {+} m_{\rm crit}  
 {+} i\gamma_5\tau^3\mu \right]\chi(x)\,,  
   \end{equation}  
where $D_W$ is the Wilson Dirac operator. For the gluon sector we use the tree-level Symanzik improved  
gauge action, $S_g$~\cite{Weisz:1982zw}. The quark fields $\chi$  
are in the so-called ``twisted basis''   
obtained from the ``physical basis''  
at  
maximal twist by a simple transformation:  
%%%TOM: \bf 1 -> \eins, so it is consistent with eq (7)  
\be  
\psi {=} \frac{1}{\sqrt{2}}[\eins + i\tau^3\gamma_5]\chi \quad {\rm and}  
\quad \bar\psi {=} \bar\chi \frac{1}{\sqrt{2}}[\eins + i\tau^3\gamma_5]\,.  
\ee  
We note that, in the continuum, this action is equivalent to the standard  
QCD action. A  
crucial advantage is the fact that by tuning a single parameter,  
%%%TOM: m_cr -> m_crit (as in eq. (1))  
namely the bare untwisted quark mass to its critical value $m_{\rm  
  crit}$,   
a wide class of physical observables are automatically ${\cal O}(a)$  
improved. A disadvantage is  the explicit flavor symmetry breaking. In  
a recent paper we have checked that this breaking is small for the baryon  
observables under consideration in this work and for the lattice spacings  
that we use~\cite{Alexandrou:2009xk,Alexandrou:2009qu,Drach:2009dh,Alexandrou:2008tn,Alexandrou:2007qq}.   
Simulations including a dynamical strange quark are also available within the  
twisted mass formulation. Comparison of the nucleon mass obtained with two  
dynamical flavors and the nucleon mass including a dynamical strange quark  
has shown negligible dependence on the dynamical strange quark~\cite{Drach:2010}.   
We therefore expect the results on the nucleon form factors to show little  
%%%TOM: sensitivity on -> sensitivity to  
sensitivity to  
a dynamical strange quark as well.

In this work we consider simulations at three values of the coupling constant  
spanning lattice spacings from about 0.05~fm to 0.09~fm. This enables us to  
examine the continuum limit of the electromagnetic form factors.  
 We find that cut-off effects are  
small for this range of lattice spacings. We also examine finite size effects  
 by comparing results on two lattices of spatial length $L=2.1$~fm and $L=2.8$~fm~\cite{Alexandrou:2010,Alexandrou:2009ng,Alexandrou:2008rp}.   
  
\section{Lattice evaluation}   
  
\subsection{Correlation functions}  
To  
extract the nucleon form factors  we need to evaluate the nucleon matrix  
element $\langle N(p^\prime,s^\prime) | {j_\mu} | N(p,s) \rangle$, where  
$|N(p^\prime,s^\prime)\rangle$, $|N(p,s)\rangle$ are nucleon states with  
%%%TOM: rewrote the sentence to avoid confusing s(s')  
final momentum $p^\prime$ and spin $s^\prime$, and initial momentum  
$p$ and spin $s$.  
  
The nucleon electromagnetic matrix element   
for real or  
virtual photons can be written in the form  
%%%TOM: these eq. are in Minkowski space? Maybe we should mention it here  
\beq  
 \langle \; N (p',s') \; | j^\mu | \; N (p,s) \rangle &=& \nonumber \\    
&\>& \hspace*{-4cm}  \biggl(\frac{ m_N^2}{E_{N}({\bf p}^\prime)\;E_N({\bf p})}\biggr)^{1/2}   
  \bar{u} (p',s') {\cal O}^{\mu} u(p,s) \; ,  
\label{NjN}  
\eeq  
where $q^2=(p^\prime-p)^2$,  
%%%TOM: energy remained unexplained  
$ m_N$ is the nucleon's mass and $E_N({\bf p})$ its energy.  
  
The operator   
${\cal O}^{\mu}$  can be decomposed in terms of the Dirac and Pauli form factors  
as  
\be  
{\cal O}^{\mu} = \gamma^\mu F_1(q^2)   
+  \frac{i\sigma^{\mu\nu}q_\nu}{2m_N} F_2(q^2) \; ,  
\label{Dirac ff}  
\ee  
%%%TOM: 1(0) was confusing  
where $F_1(0)=1$ for the proton and zero for the neutron since we have a conserved current.  
$F_2(0)$ measures the anomalous magnetic moment. They are connected to the  
electric, $G_E$, and magnetic, $G_M$, Sachs form factors by the relations  
\beq  
G_E(q^2)&=& F_1(q^2) + \frac{q^2}{(2m_N)^2} F_2(q^2)\nonumber \\  
G_M(q^2)&=& F_1(q^2) + F_2(q^2) \quad .  
\label{Sachs ff}  
\eeq

%%%TOM: The -> an  
An interpolating field for the proton in the physical basis is given by  
\be  
J(x) = \epsilon^{abc} \left[u^{a \top}(x) \C\gamma_5 d^b(x)\right] u^c(x)  
\ee  
 and can be written in the twisted basis  at maximal twist as  
\be  
\tilde{J}(x) {=} {\frac{1}{\sqrt{2}}[\eins + i\gamma_5]}\epsilon^{abc} \left[ {\tilde{u}}^{a \top}(x) \C\gamma_5 \tilde{d}^b(x)\right] {\tilde{u}}^c(x).  
\ee  
%%%TOM: third component of the iso-vector only, the others rotate to axial currents  
The
third component of the isovector current is invariant under rotation
from the physical to the twisted basis.

In order to increase the overlap with the proton state and  
decrease overlap with excited states we use Gaussian smeared quark  
fields~\cite{Alexandrou:1992ti,Gusken:1989} for the construction of  
the interpolating fields:  
%%%TOM: \vec x -> {\bf x} to be consistent with eq. (3)  
\beq  
q_{\rm smear}^a(t,{\bf x}) &=& \sum_{\bf y} F^{ab}({\bf x},{\bf y};U(t))\ q^b(t,{\bf y})\,,\\  
F &=& (\eins + {\alpha} H)^{n} \,, \nonumber\\  
H({\bf x},{\bf y}; U(t)) &=& \sum_{i=1}^3[U_i(x) \delta_{x,y-\hat\imath} + U_i^\dagger(x-\hat\imath) \delta_{x,y+\hat\imath}]\,. \nonumber  
\eeq  
In addition, we apply APE-smearing to the gauge fields $U_\mu$ entering   
the hopping matrix $H$.  
The smearing parameters   are the same as those used for our calculation of baryon masses  
with $\alpha$ and $n$ optimized for the nucleon  
ground state~\cite{Alexandrou:2008tn}. The values are: $\alpha=4.0$ and $n=50$, $70$ and $90$   
for $\beta=3.9$, $4.05$ and $4.2$ respectively.  
  
In order to calculate the  nucleon matrix element of Eq.~(\ref{NjN})   
we calculate   
the two-point and three-point functions defined by  
%%%TOM: again \vec -> \bf for consistency  
\hspace{-0.55cm}  
\beq  
\hspace{-0.55cm}G({\bf q}, t_f)\hspace{-0.15cm}&=&\hspace{-0.25cm}\sum_{{\bf x}_f} \, e^{-i{\bf x}_f \cdot {\bf q}}\,   
     {\Gamma_0^{\beta\alpha}}\, \langle {J_{\alpha}(t_f,{\bf x}_f)}{\overline{J}_{\beta}(t_i,{\bf x}_i)} \rangle \\  
\hspace{-0.5cm}  
G^\mu(\Gamma_\nu,{\bf q}, t) \hspace{-0.15cm}&=&\hspace{-0.25cm}\sum_{{\bf x}, {\bf x}_f} \, e^{i{\bf x}  
  \cdot {\bf q}}\,  \Gamma_\nu^{\beta\alpha}\, \langle  
{J_{\alpha}(t_f,{\bf x}_f)} j^\mu(t,{\bf x}) {\overline{J}_{\beta}(t_i,{\bf x}_i)}\rangle,  
\eeq  
where ${\Gamma_0}$ and ${\Gamma_k}$ are the projection matrices:  
\be  
{\Gamma_0} = \frac{1}{4}(\eins + \gamma_0)\,,\quad {\Gamma_k} =  
i{\Gamma_0} \gamma_5 \gamma_k\,.  
\ee  
\begin{figure}  
%%%TOM: this graph also needs \vec -> \bf  
 \includegraphics[width=\linewidth]{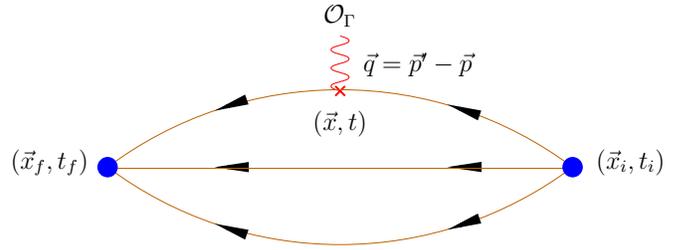}  
\caption{Connected nucleon three-point function.}  
\label{fig:connected_diagram}  
\end{figure}  
%%%TOM: I suggest to replace p and p' by p_i and p_f everywhere  
  
%%%TOM: I rephrased the following sentences and changed \vec->\bf  
The kinematical setup that we used is illustrated in  
Fig.~\ref{fig:connected_diagram}: The creation (source) operator at  
time $t_i{=}0$ has fixed spatial position ${\bf x}_i {=} {\bf 0}$. The annihilation  
(sink) operator at a later time  
$t_f$ carries momentum ${\bf p}^\prime {=} 0$. The current couples to  
a quark at an intermediate time $t$ and carries the momentum ${\bf q}$.  
Translation invariance enforces ${\bf q}=-{\bf p}$  
for our kinematics.  
The form factors are calculated as a function of  
 $Q^2=-q^2>0$, which is the Euclidean momentum transfer squared.  
%%%TOM: The correlation functions might be Euclidean already, especially eq. (11)  
%%%     maybe move the "From now on" sentence a paragraph up?  
Provided the Euclidean times, $t-t_i$ and $t_f - t_i$ are large  
enough to filter the nucleon ground state,  
the time dependence  
of the Euclidean time evolution  
%%%TOM: for -> in  
and the overlap factors cancel in the ratio  
%%%TOM: \vec -> \bf  
\be  
R^{\mu}(\Gamma,{\bf q},t)= \frac{G^\mu(\Gamma,{\bf q},t) }{G({\bf 0},  
  t_f)}\ \sqrt{\frac{G({\bf p}, t_f-t)G({\bf 0},  t)G({\bf 0},  
    t_f)}{G({\bf 0}  , t_f-t)G({\bf p},t)G({\bf p},t_f)}}\,,  
\label{ratio}  
\ee  
 yielding a time-independent value  
%%%TOM: \vec -> \bf, also changed arguments of \Pi to make eq. (15)-(17)  
%%%     simpler and consistent with eq. (21)  
\be  
\lim_{t_f-t\rightarrow \infty}\lim_{t-t_i\rightarrow \infty}R^{\mu}(\Gamma,{\bf q},t)=\Pi^\mu (\Gamma,{\bf q}) \,.  
\label{plateau}  
\ee  
We refer to the range of $t$-values where this asymptotic behavior is observed  
within our statistical precision as the plateau range.  
%%%TOM: added footnote; also: should we maybe mention already here that we take the iso-vector one?  
We use the lattice conserved electromagnetic current\footnote{In the twisted mass formulation  
both the iso-singlet and the third component of the iso-vector vector currents are  
conserved.},   $j^\mu (x)$,  
symmetrized on site $x$ by taking  
\be  
j^\mu (x) \rightarrow \frac{1}{2}\left[ j^\mu (x) + j^\mu (x - \hat \mu) \right]  
\label{lattice current}  
\ee  
We can extract the two Sachs form factors from the ratio of Eq.~(\ref{ratio}) by choosing  
appropriate combinations of the direction $\mu$ of the electromagnetic current  
  and projection matrices $\Gamma$.

 Inclusion of  a complete set of hadronic
 states in the two- and three-point functions  
leads to the following expressions, written in Euclidean time:  
%%%TOM: adjusted l.h.sides; changed 4->0 to be consistent with previous eqs.  
\be \Pi^{\mu=i} (\Gamma_k, {\bf q})  = C \frac{1}{2  
m_N} \epsilon_{ijk} \; q_j \; G_M (Q^2)   
\label{GM}  
\ee  
\be \Pi^{\mu=i} (\Gamma_0, {\bf q})  = C  
\frac{q_i}{2 m_N} \; G_E (Q^2)   
\label{GE123}  
\ee  
\be \Pi^{\mu=0} (\Gamma_0, {\bf q})  = C  
\frac{E_N +m_N}{2 m_N} \; G_E (Q^2)  \; ,  
\label{GE4}  
\ee  
%%%TOM: remark: that factor also contains traces from the 2pt functions in the ratio  
%%%     which have nothing to do with the state-normalization  
where $C=  
\sqrt{\frac{2 m_N^2}{E_N(E_N + m_N)}}$ is a kinematical factor connected to the
normalization of the lattice states and the two-point functions entering in the 
ratio of Eq.~(\ref{ratio})~\cite{Alexandrou:2006ru}.  
The first observation regarding these expressions is that the  
polarized matrix element given in Eq.~(\ref{GM}), from which  
 the magnetic form factor is determined,  
does not contribute for all momenta ${\bf q}$.  
New inversions are necessary every time a different choice of the projection  
matrix $\Gamma_\alpha$ is made
and therefore to get the other components we would need two additional inversions.  
Alternatively, one can  
%%%TOM: rephrased the following sentence to explain waht we mean with optimal; also chenged S_m->Pi^comb  
construct  
a suitable linear  
combination for the nucleon sink that leads to~\cite{Alexandrou:2006ru}  
\beq  \Pi_{\rm opt}^{\mu=i}({\bf q})= \sum_{k=1}^3\Pi^{\mu=i}(  
\Gamma_k, {\bf q}) =  \frac{C}{2m_N} \biggl\{ (q_3-q_2)\delta_{1,i} \nonumber \hspace{-0.5cm} \\  
\hspace*{-0.5cm} + (q_1-q_3)\delta_{2,i} + (q_2-q_1)\delta_{3,i} \biggr\}  
G_M(Q^2)  
\label{GM optimal}  
\eeq   
which is optimal in the sense that it provides the maximal set of  
lattice measurements from which  $G_M$  
%%%TOM: inversion -> set of inversions (which are 24 in TMQCD)  
can be extracted, requiring one set of sequential inversions.
One can choose  the sink of Eq.~(\ref{GM optimal})  
or do three inversions one for each spatial $\Gamma_i$.  
Which choice is more cost effective needs to be determined by comparing the  
statistical error at fixed cost. For the evaluation of the  
 electromagnetic form factors the two options are almost equivalent.  
 No such  
improvement is necessary for the unpolarized matrix elements  
given in Eqs.~(\ref{GE123}) and (\ref{GE4}),   
which yield $G_E$ with an additional set of sequential inversions.  
Since in this work, we consider  the temporal and spatial  
$\Gamma$'s we need a total of four sets of sequential inversions.  

%We compare the statistical error of these two options in Fig.~\ref{fig:error}.   
  
%{\bf can Pierre-Antoine provide a figure for this comparison?}  
  
%%%TOM: photon->vector current, but feel free to change it back  
 The nucleon matrix element also  contains isoscalar vector current  
contributions. This means that disconnected loop diagrams also  
contribute. These are generally difficult to  
evaluate accurately, since the all-to-all quark  
%%%TOM: added: noise  
propagator is required and the signal to noise ratio is extremely low.   
In order to avoid  
disconnected diagrams, we calculate the isovector form factors.  
 Assuming $SU(2)$ isospin  
symmetry, which holds to ${\cal O}(a^2)$ in the twisted  
mass formulation,  it follows that  
\small   
\beq   
\langle \; p \,| (  
\frac{2}{3}\bar{u} \gamma^{\mu}u - \frac{1}{3}\bar{d} \gamma^{\mu}  
d ) | p \rangle  - \langle \; n | ( \frac{2}{3}\bar{u}  
\gamma^{\mu}u - \frac{1}{3}\bar{d} \gamma^{\mu} d ) | n \rangle \;  
\nonumber \hspace{-0.9cm}\\ = \langle \; p \, | ( \bar{u}  
\gamma^{\mu} u - \bar{d} \gamma^{\mu} d ) | p \rangle .   
\eeq  
\normalsize   
One can therefore calculate directly the three-point  
function related to the right hand side of the above relation which  
provides the {\it isovector } nucleon form  
factors   
\beq   
G_E^{p-n} (Q^2) &=& G^p_E (Q^2)\, - G^n_E  
(Q^2) , \nonumber \\  
 G_M^{p-n} (Q^2) &=& G^p_M(Q^2)- G^n_M (Q^2) .  
\label{isovector}  
 \eeq  
%%%TOM: removed 'therefore', added 'directly'   
The isovector electric form factor, $G_E^{p-n}$,  
 can be obtained directly from the connected  
diagram shown in Fig.~\ref{fig:connected_diagram}.  
To extract this quantity
% the isovector  
%electric form factor $G_E^{p-n}$,  
 we consider either the spatial  
components of the electromagnetic current as given in  
Eq.~(\ref{GE123}) or  
 the temporal component given in Eq.~(\ref{GE4}).   
The isovector magnetic form factor, $G_M^{p-n}$   
is extracted using  Eq.~(\ref{GM}) for all three spatial components.
%which despite being more expensive yields errors that are  
%smaller by more than a factor of $\sqrt{3}$.  
%%%TOM: remark: this contradicts the earlier statement that  
%%%     the 2 options are almost equally expensive  

%%%TOM: take into account simultanously in our analysis all... ->  
%%%     take all... into account in our analysis  
Besides using an optimal nucleon source, the other  
important ingredient in the extraction of the form factors  
 is to take all the lattice momentum vectors that contribute to a given  
$Q^2$   into  
account in our analysis.  
  
%%%TOM: our types separate GE from GM, so no SVD is necessary  
%%%     we still can do that (I do), but it's equivalent to a  
%%%     simple weighted average. I first edited the following paragraph  
%%%     but then decided to comment it out and replace by the following:  
  
If a form factor $G(Q^2)$ can be extracted according to Eq.~(\ref{GM})-(\ref{GE4})  
from a total of $M$ directions $\mu$ and lattice momenta ${\bf q}$ and we  
denote the plateau values by $\Pi_k$, their statistical errors by $w_k$  
and the corresponding coefficient by $C_k$, the form factor is calculated  
by minimizing  
\be  
\chi^2=\sum_{k=1}^{M} \Biggl(\frac{C_{k}G(Q^2)-\Pi_k}{w_k}\Biggr)^2\, .  
\ee  
This is a least-squares fit to a constant and the result is the weighted  
average of the individual measurements.  
  
%This is done by solving the overcomplete set of equations  
%%%%TOM: P ->\Pi, matrix-notation, reformulation  
%\be  
%\Pi = D\, F(Q^2)   
%\label{overcomplete}  
%\ee  
%which is a matrix notation of Eq.~(\ref{GM})-(\ref{GE4}).  
%$\Pi$ is a vector of length $M$ containing all possible lattice measurements of the ratios  
%which have the same value of $Q^2$ and a non-zero coefficient.  
%$D$ is a $M\times 2$ matrix  
%containing the coefficients and  
%$F =  \left(\begin{array}{c}  G_{E} \\  
%                                    G_M \end{array}\right)$.  
%If $\Pi_k$ has a statistical error of $w_k$, we extract the form factors by  
%minimizing  
%%%%TOM: sum runs to M not N, and the inner one to 2 not 3  
%\be  
%\chi^2=\sum_{k=1}^{M} \Biggl(\frac{\sum_{j=1}^2 D_{kj}F_j-\Pi_k}{w_k}\Biggr)^2  
%\ee  
%using the singular value decomposition of $D$.  

%%%TOM: For 16 Q^2 values we have about 257 momentum directions. So the typical  
%%%     amount of vectors per Q^2 is O(10) not 'few hundred';  
Collecting contributions from all ${\bf q}$ directions improves the  
statistical precision and is moreover necessary to guarantee automatic $O(a)$-improvement  
with twisted mass fermions.  
%Given the fact  
%that one can have a few hundred lattice momentum vectors contributing in the evaluation  
%of the form factors,  the statistical precision is highly improved.   
 Phenomenologically interesting quantities  
like the r.m.s. radii and magnetic moments can thus be obtained with increased  
precision.  
  
%%%TOM: rephrased a little  
The connected diagram Fig.~\ref{fig:connected_diagram} is calculated  
%As already mentioned, only   
%the connected diagram contributes. It is  calculated  
 by performing sequential inversions through the sink yielding the  
form factors at all possible  
momentum transfers and current orientations $\mu$.  
Since we use a sequential inversion through the sink we need to fix the  
sink-source separation. Statistical errors increase rapidly as we increase  
the sink-source separation. Therefore we need to choose the smallest   
possible that still ensures that the nucleon ground state dominates  
%%%TOM: rephrased a little  
when measurements are made at values of $t$ in the plateau region.  
%when measurements are made at different values of $t$.  
 In order to check that a sink-source  
time separation of $\sim 1$~fm is sufficient for the isolation of the   
%%%TOM: t_f -> t_f-t_i  
nucleon ground state we compare the results at $\beta{=}3.9$   
obtained with $(t_f-t_i)/a{=}12$ i.e. $(t_f-t_i)\sim 1$~fm  
with those obtained when we increase to  
 $(t_f-t_i)/a{=}14$~\cite{Alexandrou:2008rp, Alexandrou:2010hf}.   
It was demonstrated  
that  
the plateau values for these two time separation are compatible  
yielding  the  
same results. This   
means  that the shorter sink-source separation is sufficient and the  
 ground state of the nucleon dominates in the plateau region. We therefore use in all of  
our analysis   $t_f-t_i\sim 1$~fm.  
%%%TOM: remark: the stat. errors of the t_f=14 calculation were a lot larger. Strictly speaking  
%%%     we have verified that excited state contributions are smaller than these large errors,   
%%%     not the small ones we have now. I propose to attenuate the statements slightly  
%\be  
%\Pi^{i}(\Gamma^k,\vec q){=}\frac{ic}{4m}\left[\frac{q_k q_i}{2m}\ G_p(Q^2){-}(E{+}m)\delta_{i,k}\ G_A(Q^2)\right],  
%\label{plateau1}  
%\ee  
% where $k=1,2,3, \quad{\rm and}\quad c=\sqrt{\frac{2m^2}{E(E+m)}}$.   

\subsection{Simulation details}  
  
The input parameters of the calculation, namely $\beta$, $L/a$ and $a\mu$   
%%%TOM: remark: I think here is the first time that \beta was mentioned.  
are summarized in Table~\ref{Table:params}. The  lattice spacing $a$   
is set using the nucleon mass~\cite{Alexandrou:2010hf, Alexandrou:2009qu}.  
The pion mass values, spanning a mass range   
from 260~MeV to 470~MeV, are taken   
from Ref.~\cite{Urbach:2007}.  
%%%TOM: question: are these pion masses obtained using the a from Nucleon?  
%Dina: yes
At $m_{\pi}\approx 300$ MeV and $\beta{=}3.9$ we have simulations   
for lattices of spatial size $L{=}2.1$~fm and $L{=}2.8$~fm    
allowing to investigate finite size effects.   
Finite lattice spacing effects are studied using three sets of   
results at $\beta{=}3.9$, $\beta{=}4.05$ and $\beta{=}4.2$ for the lowest  
and largest pion mass available in this work.  
These sets of gauge ensembles allow us to estimate lattice systematics  
 in order to  
%%%TOM: axial form factors -> form factors  
produce reliable predictions for the nucleon form factors.  
\begin{widetext}  
\begin{center}  
\begin{table}[h]  
\begin{tabular}{c|llllll}  
\hline\hline  
\multicolumn{6}{c}{$\beta=3.9$, $a=0.089(1)(5)$~fm,   ${r_0/a}=5.22(2)$}\\\hline   
$24^3\times 48$, $L=2.1$~fm &$a\mu$         &   &    0.0040      &   0.0064     &  0.0085     &   0.010 \\   
                               & Stat. &  &944 &210 & 365 &477 \\   
                               &$m_\pi$~(GeV) &  & 0.3032(16) & 0.3770(9) & 0.4319(12) & 0.4675(12)\\  
                               &$m_\pi L$     &    & 3.27       & 4.06      & 4.66       & 5.04     \\  
$32^3\times 64$, $L=2.8$~fm  &$a\mu$ & 0.003 & 0.004 & & & \\  
                               & Stat. & 667  &351 & & & \\  
                               & $m_\pi$~(GeV)& 0.2600(9)   & 0.2978(6) & & &  \\  
                               & $m_\pi L$    & 3.74        & 4.28      &&& \\\hline \hline  
\multicolumn{6}{c}{ $\beta=4.05$, $a=0.070(1)(4)$~fm, ${r_0/a}=6.61(3)$ }\\  
\hline  
$32^3\times 64$, $L=2.13$~fm &$a\mu$         & 0.0030     & 0.0060     & 0.0080     & \\  
                               & Stat.   &447 &325 &  419 &\\  
                               &$m_\pi$~(GeV) & 0.2925(18) & 0.4035(18) & 0.4653(15) &  \\  
                               &$m_\pi L$     & 3.32       &   4.58     & 5.28       &       \\ \hline\hline  
\multicolumn{6}{c}{ $\beta=4.2$, $a=0.056(1)(4)$~fm  ${r_0/a}=8.31$}\\\hline  
$32^3\times 64$, $L=2.39$~fm &$a\mu$          & 0.0065     &     & \\  
                               & Stat.       & 357 &  & & \\  
                               &$m_\pi$~(GeV) & 0.4698(18) &  &  \\  
                               &$m_\pi L$     & 4.24       &      & \\  
$48^3\times 96$, $L=2.39$~fm &$a\mu$         & 0.002      &      &     & \\  
                               & Stat.      & 245          &  &  & & \\  
                               &$m_\pi$~(GeV) & 0.2622(11) &  &  &  \\  
                               &$m_\pi L$     & 3.55       &  &      &       \\ \hline  
  
\end{tabular}  
\caption{Input parameters ($\beta,L,a\mu$) of our lattice calculation and corresponding lattice spacing ($a$) and pion mass ($m_{\pi}$).}  
\label{Table:params}  
\vspace*{-.0cm}  
\end{table}   
\end{center}  
  
\end{widetext}

\subsection{Determination of the lattice spacing}  
Since all quantities calculated in lattice QCD  
 are dimensionless we need to determine a scale  
to convert to physical units.  
The nucleon mass has been computed on the same  
ensembles that are now  
used here for the computation of the nucleon electromagnetic  
form factors~\cite{Alexandrou:2008tn}.  
%%%TOM: the author-list of ^ is slightly different than this one,   
%%%     so I rephrased the next paragraph slightly to refer to the  
%%%     older paper in third person  
The authors found that cut-off effects on the nucleon masses are small enough  
to justify the application of continuum chiral perturbation theory.  
Doing so the scale has been set through the nucleon mass at the  
physical point resulting in the following values:  
%Therefore, we can use the nucleon mass at the physical point  
%to set the scale.  Cut-off effects are negligible for the  
%three values of the lattice spacing considered here and we can  
%therefore use continuum chiral perturbation theory to extrapolate to  
%the physical point.  
% We obtain  
\beq  
a_{\beta=3.9}&=&0.089(1)(5)\,, \nonumber\\  
a_{\beta=4.05}&=&0.070(1)(4)\,,\nonumber\\  
a_{\beta=4.2}&=&0.056(2)(3)\,.\nonumber  
\eeq  
For a more detailed description,  
 see Refs.~\cite{Alexandrou:2010hf,Alexandrou:2008tn}.  
 The mean values are systematically higher than the lattice spacings determined from  
$f_\pi$~\cite{Baron:2009wt}, but agree within one standard deviation.
Since we are dealing with baryon properties we will use the values determined
from the nucleon mass 
%%%TOM: converting -> the conversion  
for the conversion of our lattice results to physical units.  
 We note that results on the nucleon mass using twisted mass fermions agree with  
those obtained using other ${\cal O}(a^2)$ improved formulations  
%%%TOM: 'and below' inserted  
for lattice spacings of about 0.1~fm and below~\cite{Alexandrou:2009qu}.

\section{Results}  
In this section we discuss the results obtained for the isovector electromagnetic form factors $G_E^{p-n}(Q^2)$ and $G_M^{p-n}(Q^2)$, as well as the anomalous  
magnetic moment and Dirac and Pauli mean squared radii  
derived from these form factors.  
Before extracting values that can be compared to experiment we must examine  
the volume and lattice spacing dependence of these form factors.  
As already mentioned, cut-off and volume effects were studied for the   
nucleon mass and taken into account in determining the lattice spacing.  
We perform a similar analysis in the case of the form factors.  
  
\subsection{Volume dependence}  
\begin{figure}  
\includegraphics[width=\linewidth]{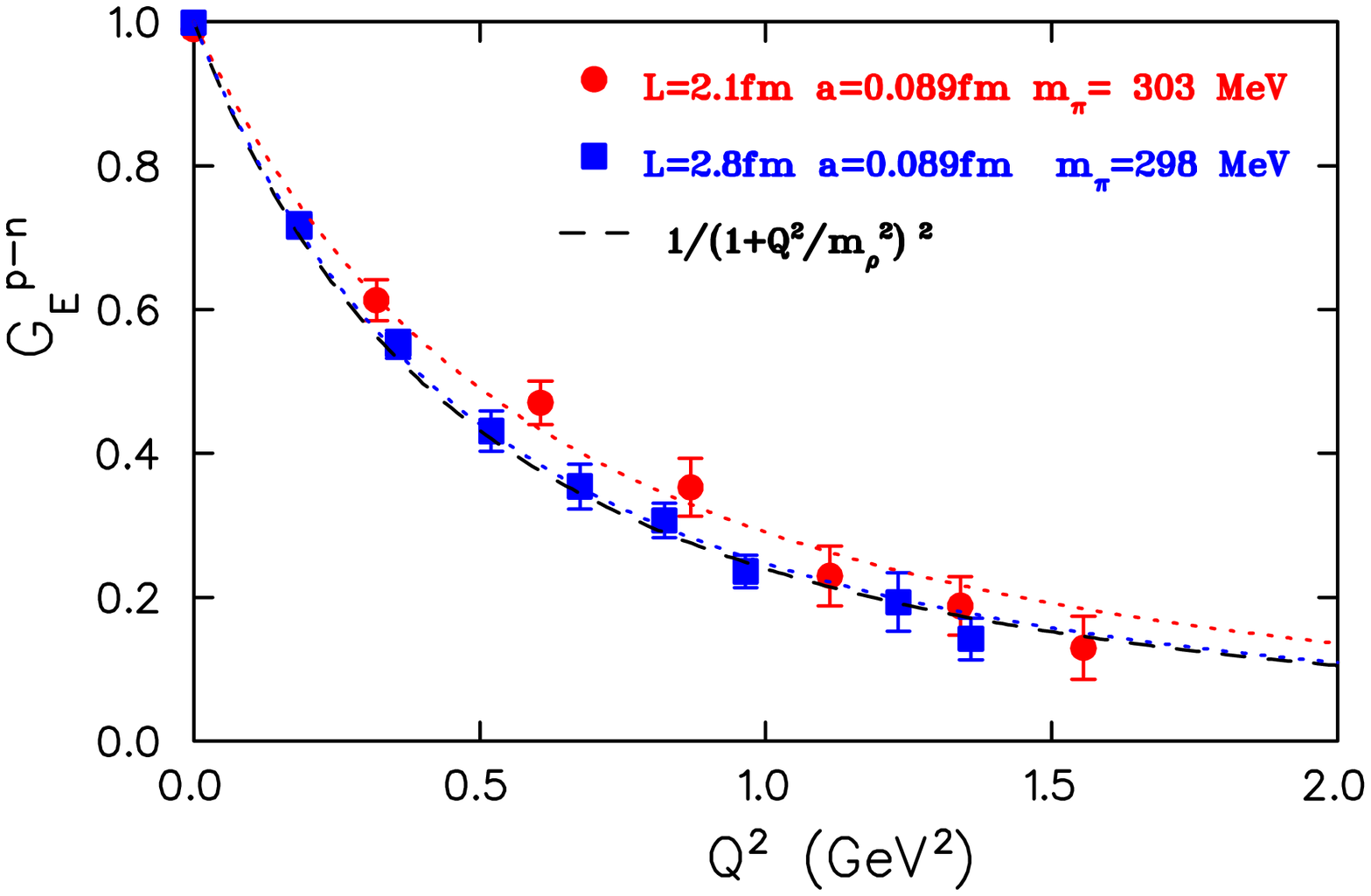}  
\includegraphics[width=\linewidth]{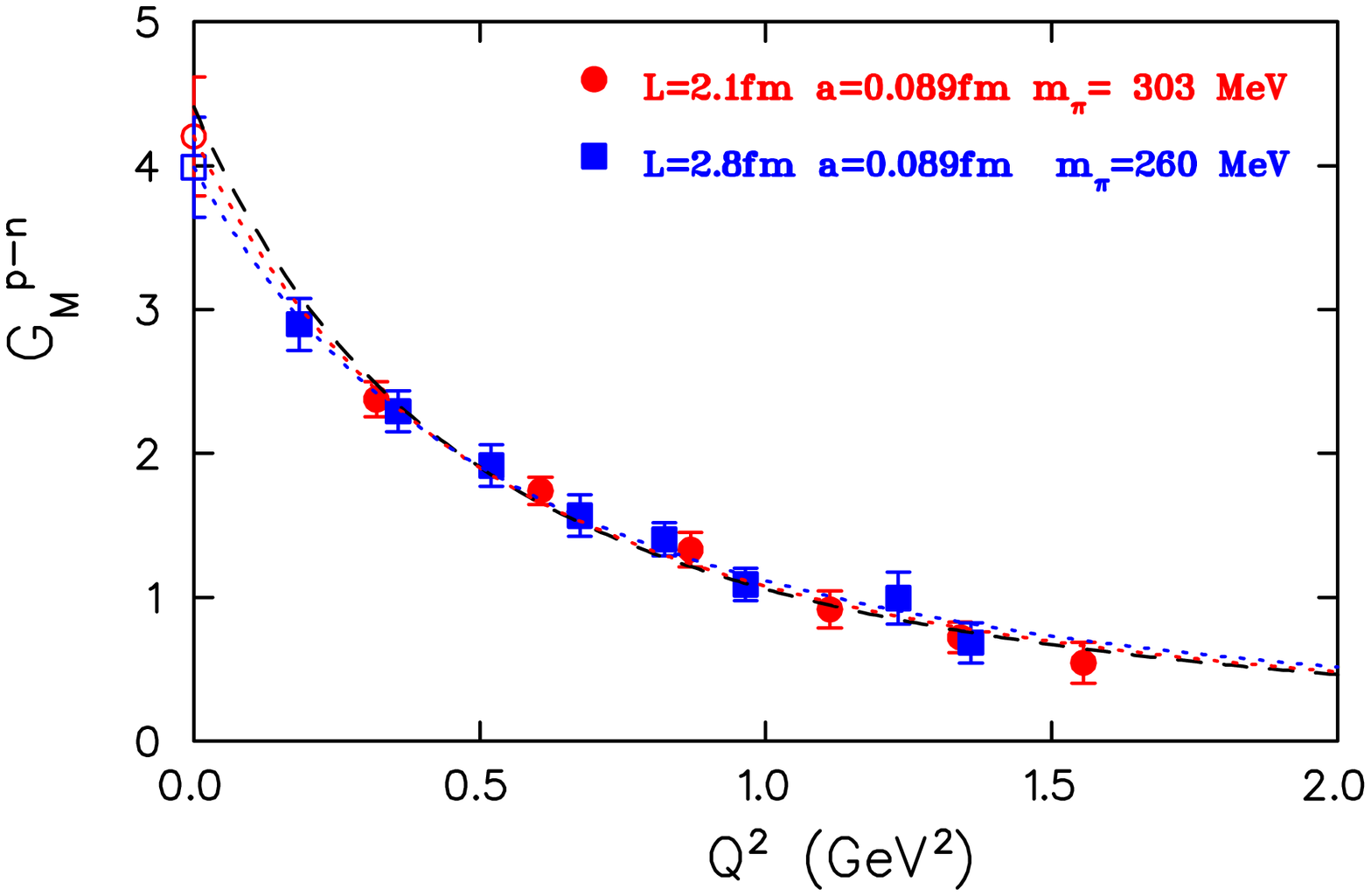}  
\caption{The nucleon isovector form factors $G_E^{p-n}$ and $G_M^{p-n}$ at $m_\pi\sim 300$~MeV  
for a lattice of size $24^3\times 48$ (filled red circles) and   
$32^3\times 64$ (filled blue squares).   
The dashed  lines correspond to  a dipole parametrization of Eq.~(\ref{dipole}) with the dipole mass  $m_E$ and $m_M$ taken to be the $\rho-$meson mass
$m_\rho$ determined on the $24^3\time 48$ lattice. The dotted lines
are dipole fits to the lattice data. 
 The value of the magnetic form factor at  
$Q^2=0$ is fitted to the lattice data.   
}  
%%%TOM: question: is the m_rho identical for both volumes? Which one is plotted?  
%%%     Maybe: mention that m_rho is the meson mass (it's explained much later in the text)  
\label{fig:GEGM L}  
\end{figure}  
In Fig.~\ref{fig:GEGM L} we check for finite volume effects by comparing results obtained at $\beta=3.9$   
on a lattice of spatial length $L=2.8$~fm and $L=2.1$~fm at  $m_\pi \sim 300$~MeV or for $Lm_\pi=3.3$ and $Lm_\pi=4.3$,  
respectively.  
%%%TOM: For GE the smaller volume data has less curvature and deviates a bit, maybe it's better  
%%%%    to replace the rather strong statement by the slightly weaker:  
As can be seen, data from both volumes are compatible with each other for $G_M^{p-n}$, indicating that  
finite volume effects are are negligible.  For $G_E^{p-n}$ there is an indication that the slope increases for the larger volume as can be seen by
the dotted lines that are dipole fits to the form
\beq  
G_E^{p-n}(Q^2){=}\frac{1}{(1{+}Q^2/m_E^2)^2}\,, \nonumber \\  
 G_M^{p-n}(Q^2){=}\frac{g_0}{(1{+}Q^2/m_M^2)^2}\,.  
\label{dipole}  
\eeq  
It is interesting to note that setting $m_E$ and $m_M$  in Eq.~\ref{dipole} to the mass of the $\rho$-meson  
as  calculated in the lattice simulations yields a very good description  
to the lattice data for the larger volume. This can be seen in Fig.~\ref{fig:GEGM L}, where the dashed line showing the dipole with the $\rho$-meson mass coincides with the dotted (blue) line, which is the fit to the data of the larger volume. The $\rho$-meson mass used is the one computed on the $24^3\time 48$ 
lattice which is in agreement with the one computed on the larger lattice. 
%As can be seen,  
%volume effects are negligible for both $G_E$ and $G_M$ for $Lm_\pi \stackrel{<}{\sim} 3.3$.  
  
\subsection{Cut-off effects}  
%%%TOM: in figure captions: asterisks -> stars   
\begin{figure}  
\includegraphics[width=\linewidth]{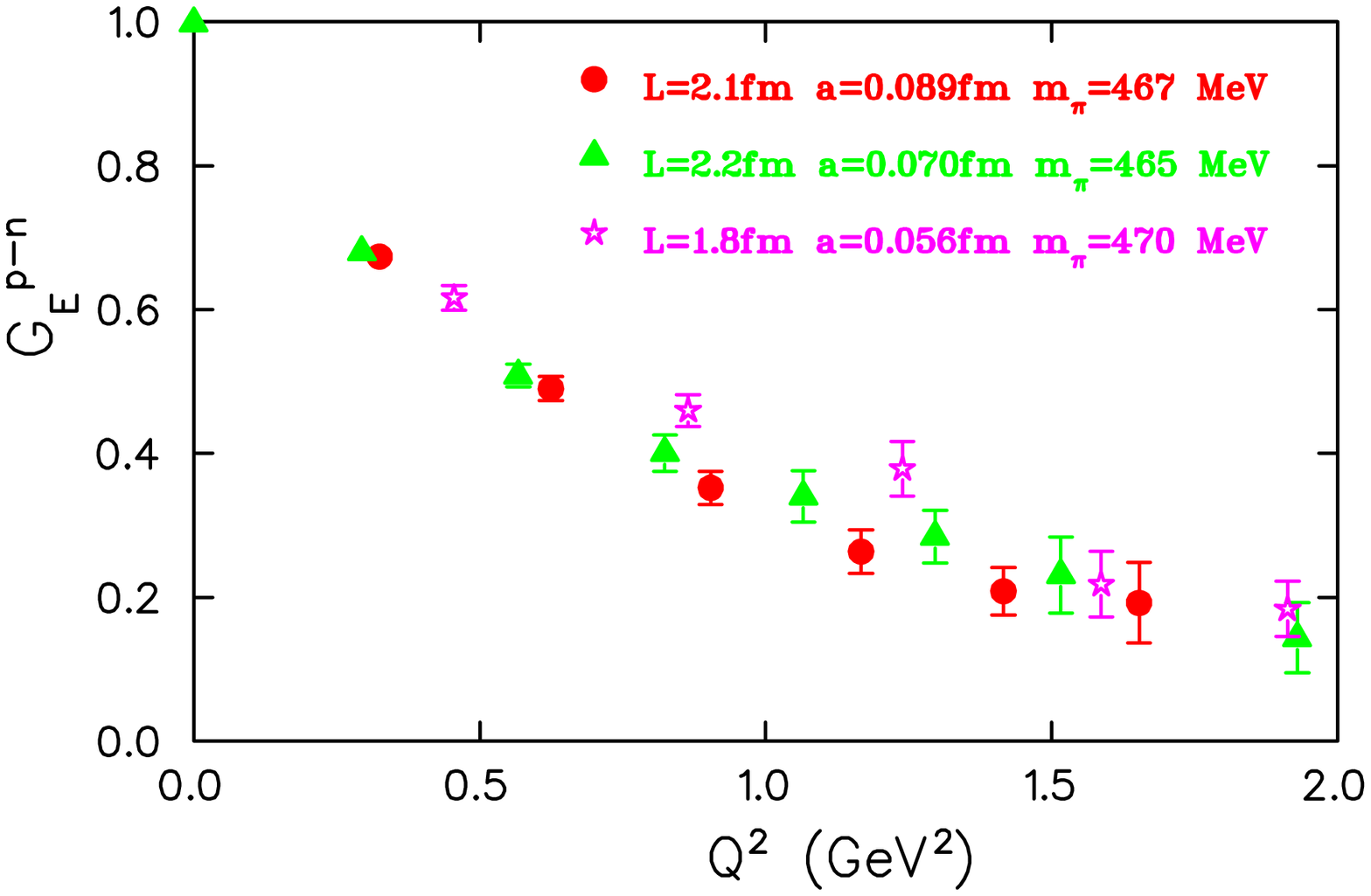}  
\includegraphics[width=\linewidth]{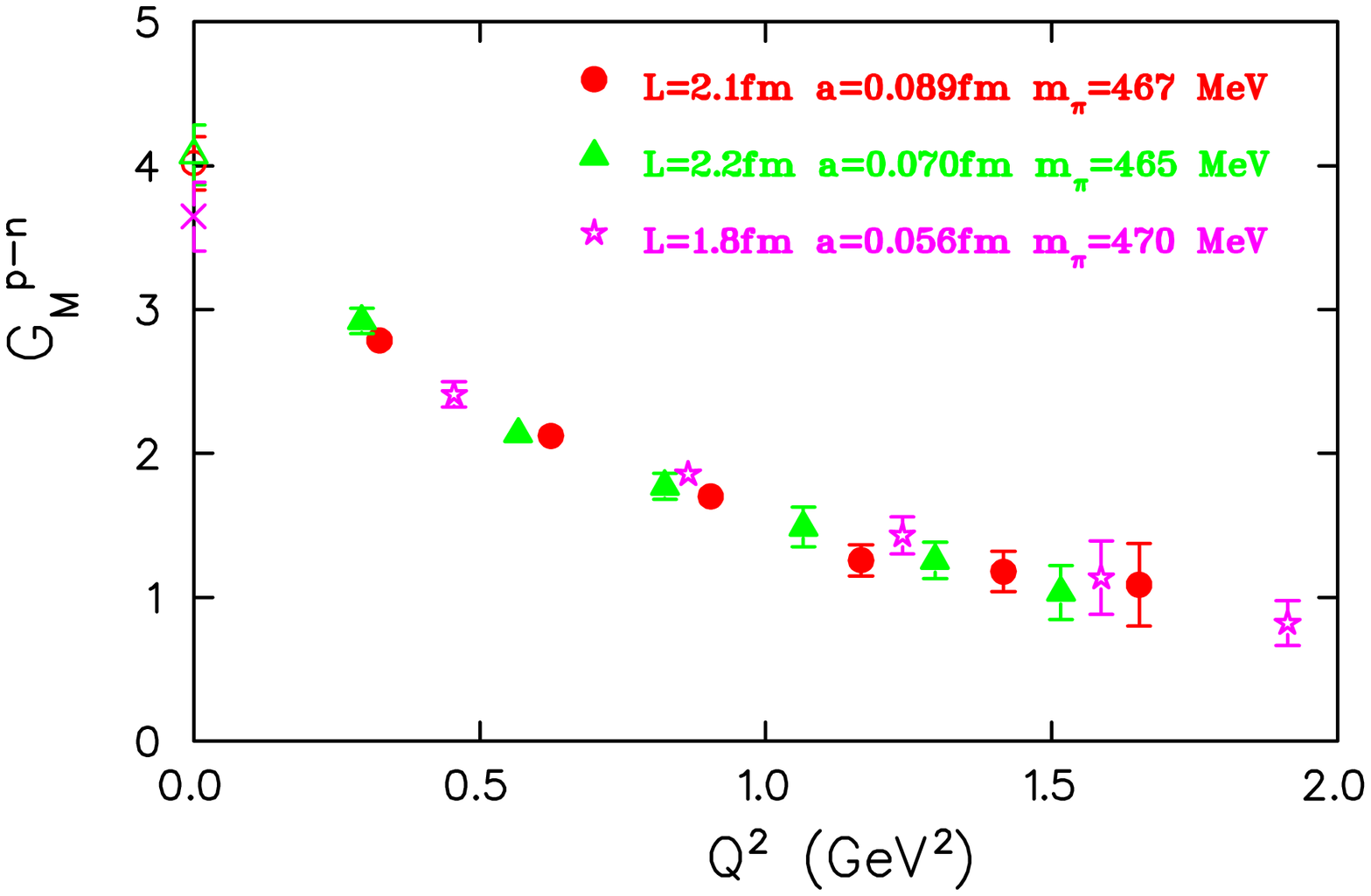}  
\caption{The nucleon isovector electric (upper) and magnetic (lower) form factors $G_E^{p-n}(Q^2)$ and $G_M^{p-n}(Q^2)$ at $m_\pi\sim 470$~MeV  
at $\beta=3.9$ (filled red circles), $4.05$ (filled green triangles) and $4.2$ (magenta stars) versus $Q^2$. The open symbols and crosses denoted the values at $Q^2=0$ at $\beta=3.9, 4.05,$ and $4.2$ respectively,  
extracted by fitting the data to a dipole form.   
}  
\label{fig:GEGM a}  
\end{figure}  
\begin{figure}  
\includegraphics[width=\linewidth]{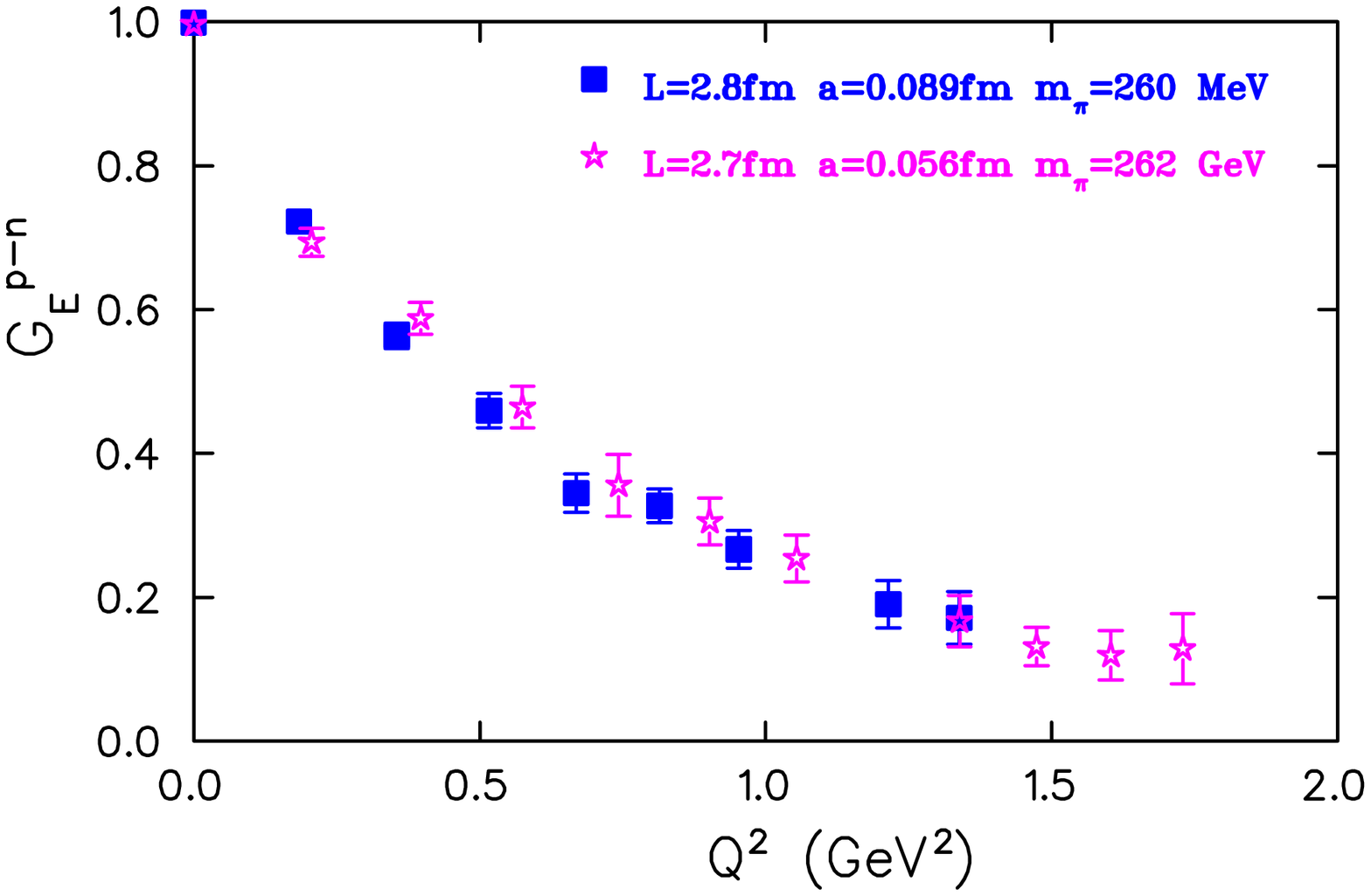}  
\includegraphics[width=\linewidth]{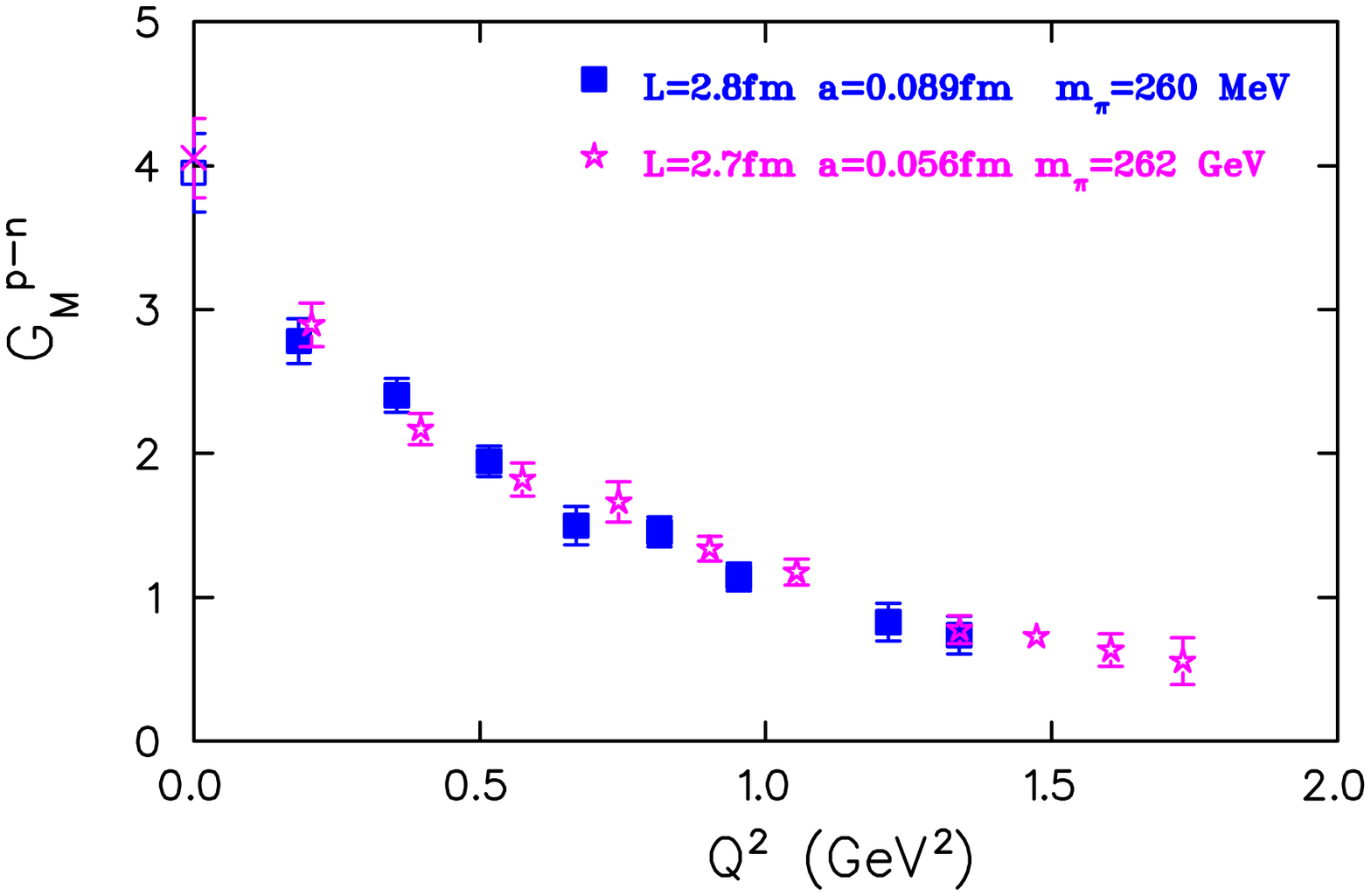}  
\caption{The nucleon isovector electric (upper) and magnetic (lower) form factors $G_E^{p-n}(Q^2)$ and $G_M^{p-n}(Q^2)$ at $m_\pi\sim 260$~MeV  
at $\beta=3.9$ (filled red circles) and $4.2$ (magenta stars) versus $Q^2$. The open symbols and crosses denoted the values at $Q^2=0$ at $\beta=3.9$ and $4.2$ respectively,  
extracted by fitting the data to a dipole form.  
}  
\label{fig:GEGM2 a}  
\end{figure}  
  
To assess cut-off effects we compare in Figs.~\ref{fig:GEGM a} and  
\ref{fig:GEGM2 a} results for  
 $G_E^{p-n}(Q^2)$ and $G_M^{p-n}(Q^2)$   
for three different lattice spacings at a similar pion mass. We consider  
results at our  
heaviest and lightest pion masses.  
For $G_M^{p-n}(Q^2)$, results at these three lattice spacings  are   
consistent for both heavy and light mass indicating  
%%%TOM: negligible -> negligible at our current statistical precision  
that cut-off effects are negligible for these lattice spacings at our current statistical precision.  
There is also consistency for the results obtained  
for  $G_E^{p-n}(Q^2)$ at the smaller pion mass of 260~MeV, as can be  
seen in Fig.~\ref{fig:GEGM2 a}. At the heavier pion mass of 470~MeV,   
a couple of data obtained on the finest lattice have higher values,
which however are well within the statistical fluctuations.
%%%TOM: the sentence below makes no sense: the statistical fluctuations are already reflected in the error bars.  
%%%     also: isn't the lattice-volume more important than the physical one for the errors? (Both are 24^3x48 I guess)  
%Given that this lattice is the smallest, namely 1.8~fm, one expects statistical larger fluctuations in the data.  

\subsection{Mass dependence of form factors}  
\begin{figure}  
\includegraphics[width=\linewidth]{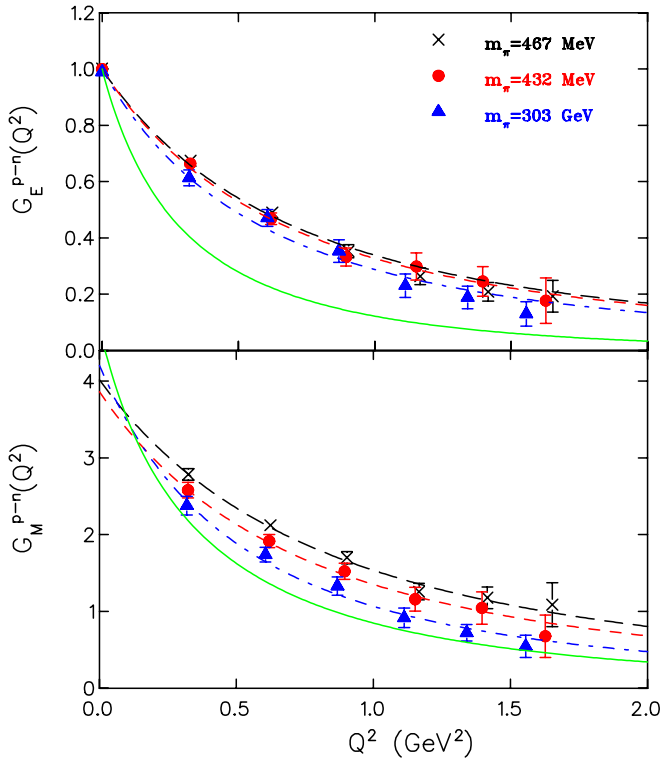}              
\caption{The nucleon  form factors $G_E^{p-n}(Q^2)$ and $G_M^{p-n}(Q^2)$   
at $\beta=3.9$ for $m_\pi = 468 $~MeV (crosses), $m_\pi = 432 $~MeV (filled red circles) and $m_\pi = 303 $~MeV (filled blue triangles) versus $Q^2$. The dashed lines are the result of a dipole fit to the lattice data.The solid line is J. Kelly's parametrization to the experimental data~\cite{Kelly:2004hm}.}  
\label{fig:GEGM mpi}  
\end{figure}

Our lattice simulations use light quark masses that correspond to pion masses  
in the range of about 470~MeV to 260~MeV. In order to obtain results  
at the physical point we need to study the dependence  on the quark mass or equivalently on the pion mass.  
We show in Fig.~\ref{fig:GEGM mpi} the dependence on the pion mass  
at a fixed volume and lattice spacing.  
We show both $G_E^{p-n}$ and $G_M^{p-n}$   
computed at several values of the pion mass  
spanning pion masses from about 470~MeV to 300~MeV at $\beta=3.9$.   
%As can be  
%seen, as the pion mass decreases the form factors  
%have a steeper $Q^2$-dependence.   
%Therefore, the lattice data tend to approach the experimental results  
%shown by the solid line and represent Kelly's parametrization  
%of the experimental data~\cite{Kelly:2004hm}.   
  
In order to extract the anomalous magnetic moment and mean squared radii we  
need to perform a fit to the $Q^2$-dependence of the form factors.  
For both $G_E^{p-n}(Q^2)$ and $G_M^{p-n}(Q^2)$ we use a dipole of the  
form given in Eq.~(\ref{dipole}). 
We fit the lattice data  
using all data up to a largest $Q^2$ value of $\sim 1.5$~GeV$^2$.  
%%%TOM: described->describe; the Kelly's-> Kelly's  
%%%     remark: maybe we should cite the dipole mass amd GM(0) of Kelly  
These fits, shown in Fig.~\ref{fig:GEGM mpi}, describe the results very  
well. On the same figures we also show Kelly's parametrization of the  
experimental data~\cite{Kelly:2004hm}.  Although  the mass dependence   
is weak and lattice data  show a weaker $Q^2$-dependence  
%%%TOM: as compared to -> than experimental data; data->results  
than experimental data, the general trend is that lattice results  approach  
experiment as the light quark mass decreases towards its physical value.  
%%%TOM: the next sentence repeats almost exactly what has already been said  
%Therefore, the lattice data tend to approach the experimental results   
%represented by  
%the solid line, which is Kelly's parametrization  
%of the experimental data~\cite{Kelly:2004hm}.    
The values extracted from the fits for the  
 dipole electric and magnetic masses, $m_E$ and $m_M$,  
 are therefore larger than in  
%%%TOM: as expected -> which is expected given the; shown->exhibited  
experiment which is expected given the smaller slope exhibited by the lattice  
data.  
As we discuss in the next section, this behavior is observed with  
%%%TOM: inserted: 'as well'  
other lattice discretization schemes as well.  
In Table~\ref{Table:GE GM fits} we tabulate the resulting fitting  
parameters for all $\beta$ and $\mu$ values. The parameters $m_E$, $G_M(0)$  
and $m_M$ have been extracted from  
fits to the form given in Eq.~(\ref{dipole}).

%%%TOM: question: is there a reference that predicts this kind of vector-meson dominance?  
  
  \begin{table}[h]  
\begin{center}  
\begin{tabular}{c|c|c|c|c|c}  
\hline\hline  
$a\mu$ & $m_\pi$ (GeV)   & $m_E$ (GeV)& $G_M(0)$ & $m_M$ (GeV) &$\mu_N$  \\\hline   
\multicolumn{6}{c}{$\beta=3.9$}\\\hline   
0.0100 & 0.4675 & 1.40(7)  & 4.02(18)   & 1.62(17)  &  2.57(22)   \\  
0.0085 & 0.4319 & 1.33(10) & 3.86(25)   & 1.45(21)  &  2.49(28)   \\  
0.0064 & 0.3770 & 1.19(9)  & 3.91(35)   & 1.69(36)  &  2.70(33)   \\  
0.004  & 0.3032 & 1.16(10) & 4.20(41)   & 1.01(14)  &  2.20(26)   \\  
0.004  & 0.2978 & 0.98(6)  & 3.99(35)   & 1.11(16)  &  2.13(18)    \\  
0.003  & 0.2600 & 1.03(6)  & 3.95(27)   & 1.13(13)  &  2.14(12)    \\\hline  
\multicolumn{6}{c}{$\beta=4.05$}\\\hline   
0.008 & 0.4653   & 1.41(74)   & 4.07(21)  & 1.56(19)  & 2.53(49)  \\  
0.006 & 0.4035   & 1.24(10)   & 4.26(30)  & 1.43(20)  & 2.41(21)   \\  
0.003 & 0.2925   & 1.31(18)   & 4.13(62)  & 1.09(27)  & 2.16(14)   \\\hline  
\multicolumn{6}{c}{$\beta=4.2$}\\\hline   
0.0065& 0.4698  &  1.75(12)   & 3.65(24) & 2.09(31)   & 2.62(24)    \\  
0.002 & 0.2622  &  1.12(8)    & 4.05(28) & 1.17(12)   & 1.84(17)    \\\hline  
\end{tabular}  
\caption{Results on the nucleon electric and magnetic mass extracted by fitting  
to a dipole form. The two last column give the $G_M(0)$ and the   
 nucleon anomalous magnetic moment in Bohr magnetons by fitting $G_M^{p-n}(Q^2)$ to the dipole form given in Eq.~(\ref{dipole}).}  
\label{Table:GE GM fits}  
\end{center}  
\end{table}

\subsection{Comparison of lattice results}  
\begin{figure}  
\includegraphics[width=\linewidth]{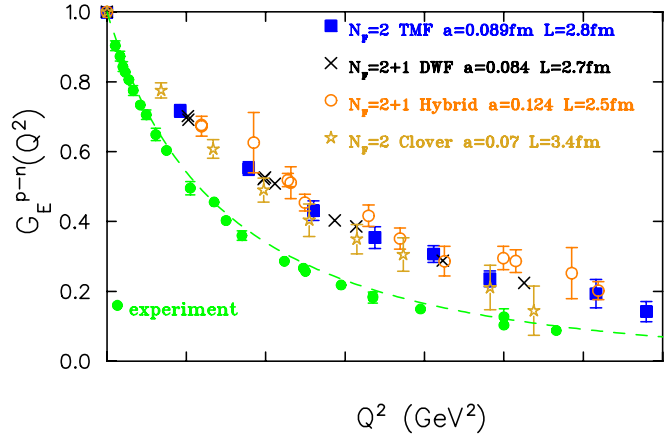}  
\caption{Isovector electric form factor $G_E^{p-n}(Q^2)$ as a function of $Q^2$. $N_F=2$ TMF  
results at $m_\pi=298$~MeV  are shown with filled squares, $N_F=2+1$ DWF~\cite{Syritsyn:2009mx}  
 (crosses), hybrid~\cite{Bratt:2010jn} (open orange circles) and Clover fermions~\cite{Capitani:2010sg} (yellow stars).   
Experimental data are shown with the filled green circles   
accompanied with Kelly's parametrization shown with the dashed line.  
}  
\label{fig:GE}  
\end{figure}

\begin{figure}  
\includegraphics[width=\linewidth]{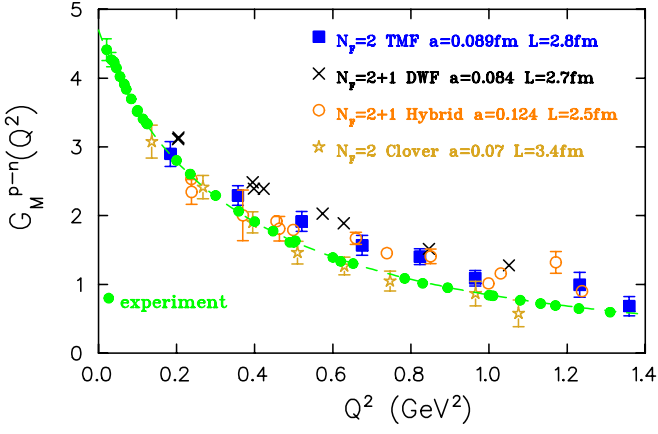}  
\caption{Isovector magnetic form factor $G_M(Q^2)$ as a function of $Q^2$. The notation  
is the same as that of Fig.~\ref{fig:GE}.  
 }  
\label{fig:GM}  
   \end{figure}

We have shown that volume and cut-off effects are small on the isovector form  
factors for the parameters used in our simulations.   
%%%TOM: Therefore this allows -> This justifies to some extent  
This justifies to some extent  
a comparison with results of other collaborations that use different  
%%%TOM: with -> but in  
fermions but in a similar volumes and at a similar lattice spacings. In Figs.~\ref{fig:GE} and  
%%%TOM: reorganized the sentence  
\ref{fig:GM} we show a comparison of the results of this work with those  
obtained using $N_F=2+1$ dynamical domain   
wall fermions (DWF)~\cite{Syritsyn:2009mx}, $N_F=2$ Wilson improved Clover  
fermions~~\cite{Capitani:2010sg} and using a hybrid action of tadpole-improved  
$N_F=2+1$ staggered fermions and domain wall valence quarks~\cite{Bratt:2010jn}   
for a pion mass around 300~MeV.  
%%%TOM: improvement -> agreement  
We can see a nice agreement among all lattice results for $G_E^{p-n}$.  
%%%TOM: the -> all; as compared to-> than  
As already pointed out, all lattice results show a weaker $Q^2$-dependence  
than experiment. 
We would like to note that in Ref.~\cite{Bratt:2010jn}  hybrid results  
for $m_\pi=356$~MeV are obtained on a lattice of $L=2.5$ and $L=3.5$~fm.
 A comparison between these results showed  volume effects for the isovector $F_1$
but not for the isovector $F_2$. This is consistent with our results on the
isovector electric and magnetic form factors 
shown in Fig.~\ref{fig:GEGM L}.
The experimental data are obtained by interpolating the neutron form factors
to the $Q^2$-values of the proton form factors as described in ref.~\cite{Alexandrou:2006ru}. 
In the case of $G_M^{p-n}$ there are discrepancies, in particular, with  the results using Clover fermions which are systematically lower.  
%%%TOM: sentence slightly rephrased  
We note that compared to $G_E^{p-n}$, $G_M^{p-n}$ is more sensitive to the  
nucleon mass which is needed as an input. Clearly one has to study further  
the systematics on the various  
lattice results, some of which are still preliminary,   
in order to clarify these discrepancies.  
  
%%%TOM: general question to figures which show experimental points:  
%%%     how were experimental values for Iso-Vector quantities obtained;   
%%%     Were Neutron and Proton experiments carried out at exactly the same   
%%%     Q^2 values?  
%Dina: I have interpolated to the same Q^2
  
\subsection{Chiral extrapolation}  
%%%TOM: remark: I didn't understand the first sentence until after I   
%%%%    read the section. But I also don't have an idea how to reformulate it at the moment.  
Given that cut-off effects are small, we use chiral  
perturbation theory that holds in the continuum to   
study the quark mass dependence of the electromagnetic form factors  
down to the physical point. This will be justified 
in the next section 
where we discuss the continuum extrapolation of our results.    
For the chiral extrapolation, we use our TMF results that  cover a  
range of pion masses from about 470~MeV down to about  
260~MeV. The pion mass dependence for the isovector form factors as well as for the anomalous magnetic moment and radii  
have been studied within  HB$\chi$PT in the so called small scale expansion (SSE) formulation~\cite{Hemmert:2002uh}.

The anomalous magnetic moment, which is given by the isovector Dirac
form factor $F_2(0)$ is extracted by fitting the $Q^2$ assuming
e.g. a dipole form dependence. 
The slope of $F_1$ at $Q^2=0$ determines the transverse size of the hadron,
$<r_\perp^2> = -4 dF_1/dQ^2|_{Q^2=0}$. In
the non-relativistic limit  the root mean square (r.m.s.) radius is related to
the slope of the form factor at zero momentum transfer. Therefore 
the r.m.s. radii can
 be obtained  from the values of the dipole masses by using
\be
<r_i^2>=-\frac{6}{F_i(Q^2)}\frac{dF_i(Q^2)}{dQ^2}|_{Q^2=0}=\frac{12}{m_i} \hspace*{0.5cm} i=1,2 \quad.
\ee
The electric and magnetic radii are 
 given by $<r_{e,m}^2>=12/m_{E,M}$ and can be directly evaluated from
the values given in Table~\ref{Table:GE GM fits}.

\begin{figure}  
\includegraphics[width=\linewidth]{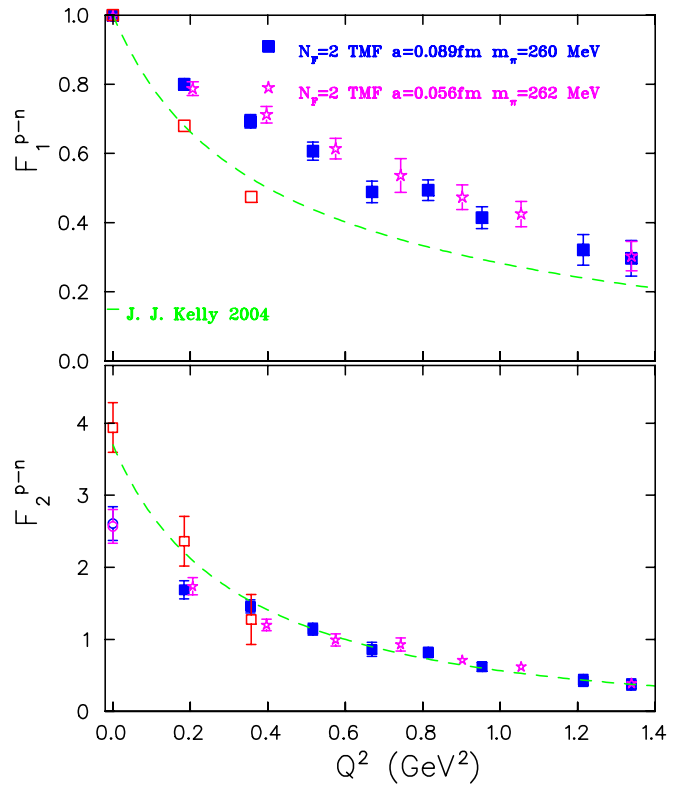}  
\caption{$F_1$ and $F_2$ for the coarser and finer lattices at  
pion mass of about 260 MeV. The dashed line is Kelly's parametrization  
to the experimental data. The open squares show the values of  
$F_1$ and $F_2$ after chiral extrapolation to the physical point.  
}  
\label{fig:F1F2}  
\end{figure}

 Using HB$\chi$PT to one-loop, with $\Delta$ degrees  
 of freedom and iso-vector  $N$-$\Delta$ coupling included in LO~\cite{Hemmert:2002uh,Gockeler:2003ay} the expression for  
the  isovector anomalous magnetic moment $\kappa^{p-n}$ is given by~\cite{Gockeler:2003ay}  
\beq       
\kappa^{p-n}(m_\pi)&=&{\ \kappa_{\rm v}(0)} -\frac{g_A^2 m_\pi m_N}{4\pi f_\pi^2}   
\nonumber \\ &&\hspace*{-2cm}  
 +\frac{2{c_A^2}\Delta m_N}{9\pi^2 f_\pi^2}\biggl [\sqrt{1-\frac{m_\pi^2}{\Delta^2}} \log R(m_\pi)   
+ \log\left (\frac{m_\pi}{2\Delta}\right)\biggr]\nonumber\\ &&\hspace*{-2cm}  
-8{E_1} m_N m_\pi^2  + \frac{4 c_A c_V g_A m_N m_\pi^2}{9\pi^2f_\pi^2}  
\log\left(\frac{2\Delta}{\lambda}\right)\nonumber \\ & & \hspace*{-2cm}  
 + \frac{4{c_A} {c_V} g_A m_N m_\pi^3}{27 \pi F_\pi^2 \Delta}  
 - \frac{8 {c_A c_V} g_A \Delta^2 m_N}{27 \pi^2 f_{\pi}^2}\nonumber \\ & & \hspace*{-2cm}    
\biggl[ \left(1-\frac{m_\pi^2}{\Delta^2}\right)^{3/2}\log R(m_\pi)   
\nonumber \\  
&+& \left(1-\frac{3m_\pi^2}{2\Delta^2}\right) \log\left(\frac{m_\pi}{2\Delta}  
\right) \biggr ]\,,  
  \eeq  
and for the isovector Dirac form factor~\cite{Gockeler:2003ay}   
\begin{widetext}  
\begin{eqnarray}   
{} & {} &   
F_{1}^{p-n}(m_\pi,Q^2) =  1 + \frac{1}{(4\pi f_\pi)^2}  
                   \left\{ -Q^2 \left(\frac{68}{81} c_A^2   
           - \frac{2}{3}g_{A}^2-2 B_{10}(\lambda) \right) \right.   
   - Q^2 \left(\frac{40}{27} c_A^2-\frac{5}{3}g_{A}^2  
   -\frac{1}{3}\right)  
                   \log\left[\frac{m_\pi}{\lambda}\right]    
\nonumber \\  {} & {} &   
     {} + \int_{0}^{1}dx \left[\frac{16}{3}\Delta^2 c_A^2  
          + m_{\pi}^2 \left(3 g_{A}^2+1-\frac{8}{3} c_A^2 \right) \right.  
        + \left. Q^2 x(1-x)\left(5 g_{A}^2+1  
        -\frac{40}{9}c_A^2\right)\right]   
        \log\left[\frac{\tilde{m}^2}{m_{\pi}^2}\right]    
\nonumber \\ {} & {} &   
     {} + \int_{0}^{1}dx \left[ -\frac{32}{9} c_A^2 Q^2 x(1-x)   
       \frac{\Delta \log R(\tilde{m})}{\sqrt{\Delta^2-\tilde{m}^2}} \right]   
\nonumber \\  {} & {} & {}  
    -  \left. \int_{0}^{1}dx \; \frac{32}{3}c_A^2 \Delta   
      \left[\sqrt{\Delta^2-m_{\pi}^2}\log R(m_\pi)   
      -\sqrt{\Delta^2-\tilde{m}^2}\log R(\tilde{m}) \right] \right\}   
      + \mathcal O (\epsilon^4) \,.   
\label{F1}  
\end{eqnarray}  
%%%TOM: 'the expansion of' inserted  
To the same order the expansion of the isovector Pauli form factor is given by  
\begin{multline}  \label{F2}  
 F_{2}^{p-n}(m_\pi,Q^2) =  \kappa^{p-n} (m_\pi) - g_{A}^2  
           \frac{4\pi m_N}{(4\pi f_\pi)^2}   
           \int_{0}^{1}dx\left[ \sqrt{\tilde{m}^2}-m_{\pi}\right]  
         {} +\frac{32c_A^2 m_N \Delta}{9 (4\pi f_\pi)^2}\int_{0}^{1}dx   
         \left[ \frac{1}{2}\log\left[\frac{\tilde{m}^2}{4\Delta^2}\right]  
         -\log\left[\frac{m_\pi}{2\Delta}\right] \right.    
\\  
  \left. {} +\frac{\sqrt{\Delta^2-\tilde{m}^2}}{\Delta} \log R(\tilde{m})  
              -\frac{\sqrt{\Delta^2-m_{\pi}^2}}{\Delta} \log R(m_\pi) \right] \,,  
\end{multline}  
\end{widetext}  
where   
\begin{equation}  
R(m)  
= \frac{\Delta}{m}+\sqrt{\frac{\Delta^2}{m^2}-1} \,, \quad  
\tilde{m}^2 = m_\pi^2 + Q^2 x (1-x) \,.  
\end{equation}  
We perform a  fit  to   
$F_1(m_\pi,Q^2)$ and $F_2(m_\pi,Q^2)$ with five parameters, namely the  
iso-vector magnetic moment at the chiral limit {$\kappa_{\rm v}(0)$},   
the isovector and axial N to $\Delta$   
coupling constants, $c_V$ and $c_A$ and the two counterterms $B_{10}(\lambda)$ and  
$E_1(\lambda)$.   
The rest of the parameters are fixed to their physical values, namely  
%%%TOM: removed 'we take'  
$m_N=0.938$~MeV, $g_A=1.267$, $f_\pi=0.0924$~MeV and the $\Delta$-nucleon mass splitting  
$\Delta=0.2711$. The counterterms are evaluated at $\lambda=0.6$~GeV.  
For  
the chiral extrapolation we use data at the three lowest $Q^2$-values.  
%%%TOM: question: is there some common wisdom up to which Q^2 these expansions work?  
% Not really. One just looks at chi^2
As can be seen in Fig.~\ref{fig:F1F2}, the chiral extrapolation   
decreases the value of $F_1$ and increases the value of $F_2$  
at low $Q^2$,  bringing them into qualitative agreement with experiment.  
 The values of the fit parameters are given in Table~\ref{tab:fit params}.  
Although such a chiral extrapolation is useful and probes  the general trend  
we would like to point out that the $\chi^2$ per degree of freedom (d.o.f)  
is 4.3, which means that the description of the results is not really optimal.  
  
%%%TOM: question to the fits: how manz d.o.f. are there in total?  
%%%     are the three Q^2 values fitted in one common fit?  
%Dina; Yes all 3 Q^2 are fitted together  
  
\begin{table}[h]  
\begin{tabular}{c|c|c}  
\hline  
Fit parameter & data at 3 $\beta$ values & continuum data   \\\hline  
\multicolumn{3}{c}{Fit separately $\kappa^{p-n}$, $r_1^{2\, p-n}$ and $r_2^{2\,p-n}$}\\\hline  
$\kappa_{\rm v}(0)$ &4.22(84) & 4.22(74)\\  
$c_V$               &-5.86(3.52) & -5.46(2.73)\\  
$E_1(0.6\,{\rm GeV})$  &-8.95(3.80) &  -8.58(3.00)\\  
$B_{10}(0.6\,{\rm GeV})$  & 0.04(2) & 0.04(2) \\  
$B_{c2}(0.6\,{\rm GeV})$  & 0.14(3) & 0.12(3) \\\hline  
\multicolumn{3}{c}{Fit to the Dirac and Pauli form factors}\\\hline  
$\kappa_{\rm v}(0)$ &5.57(56) & 5.20 (20)\\  
$c_A$               &1.56(3)  & 1.53(2)\\  
$c_V$               &-2.20(1.62) & -3.25(53)\\  
$E_1(0.6\,{\rm GeV})$  &-6.45(2.42) &  -7.87(69)\\  
$B_{10}(0.6\,{\rm GeV})$  & 1.21(6) & 1.11(6) \\  
\hline   
\end{tabular}  
\caption{The values of the parameters extracted from the fit to $F_1$ and  
$F_2$ using lattice data at $\beta=3.9$, $\beta=4.05$ and $\beta=4.2$ (second column) and using the data after taking the continuum limit (third column).}  
\label{tab:fit params}  
\end{table}

\begin{figure}  
\includegraphics[width=\linewidth]{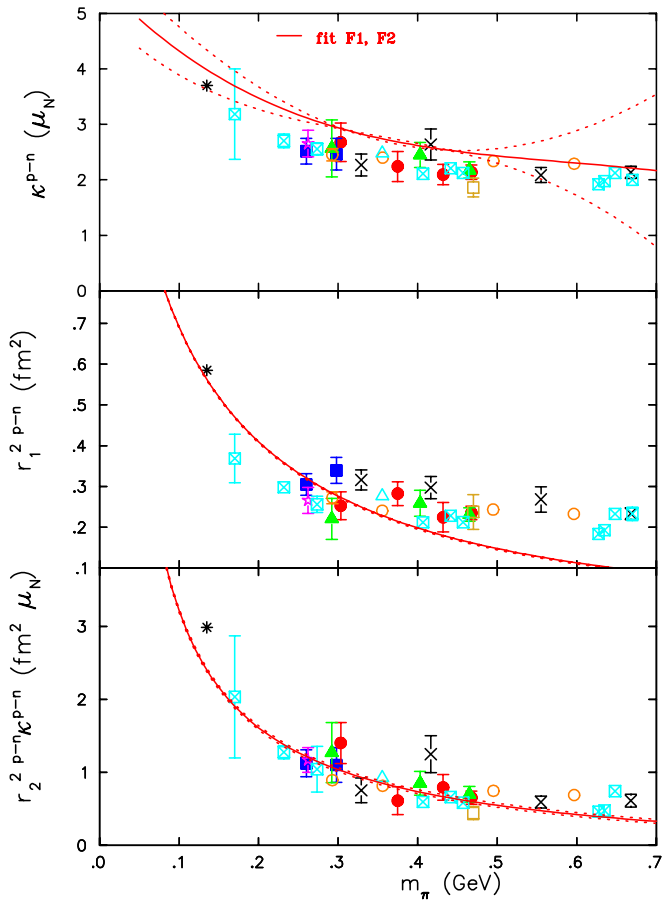}  
\caption{Chiral fits using the parameters determined from fitting $F_1$ and  
$F_2$. We show lattice data  using $N_F=2$ TMF~\cite{Alexandrou:2010hf}   
($a=0.089$~fm: filled red  
circles for $L=2.1$~fm and filled blue squares for $L=2.8$~fm;  $a=0.070$~fm: filled green triangles for $L=2.2$~fm; $a=0.056$~fm:  
purple star for $L=2.7$~fm and open yellow square for $L=1.8$~fm),  $N_F=2+1$  DWF~\cite{Yamazaki:2009zq} (crosses for $a=0.114$~fm and $L=2.7$~fm),  $N_F=2+1$ using DWF and staggered sea~\cite{Bratt:2010jn} ($a=0.124$~fm: open orange circles for $L=2.5$~fm and open cyan triangle for $L=3.5$~fm) and using $N_F=2$ Clover fermions~\cite{Collins:2011gw} are also shown with the cyan cross-in-square symbols. The physical point is marked by the asterisk.  
}  
\label{fig:kappav r1 r2}  
\end{figure}  
Using the fit parameters determined from $F_1$ and $F_2$ we can obtain  
the mass dependence of the isovector magnetic moment and radii.    
The expressions for  the radii $r_1^{2\, p-n}$  
and $r_2^{2\, p-n}$ are given by~\cite{Gockeler:2003ay}  
\beq  
r_1^2 &=& -\frac{1}{(4\pi f_\pi)^2}\biggl[1+7g_A^2+\left(10g_A^2+2\right)  
\log\left(\frac{m_\pi}{\lambda}\right)\biggr] \nonumber \\  
& -&\frac{12 B_{10}}{(4\pi f_\pi)^2}   
+ \frac{{c_A^2}}{54\pi^2 f^2_\pi}\biggl[26 + 30 \log\left(\frac{m_\pi}{\lambda}\right) \nonumber \\  
&  +&30 \frac{\Delta}{\sqrt{\Delta^2-m_\pi^2}}\log R(m_\pi) \biggr]\\  
r_2^2&=& \frac{1}{\kappa_{\rm v}(m_\pi)}\biggl\{   
\frac{g_A^2 m_N}{8f_\pi^2\pi m_\pi}   
+ \frac{c_A^2 m_N}{9f^2_\pi \pi^2\sqrt{\Delta^2-m_\pi^2}}\log R(m_\pi)\nonumber \\  
 &+&24 m_N {\ B_{c2}} \biggr\}   
\label{r1 r2}
\eeq  
% & & R_2(m)=\frac{\Delta}{2\sqrt{\Delta^2-m^2+i\epsilon}}   
%\log\left(\frac{\Delta+\sqrt{\Delta^2-m^2+i\epsilon}}{\Delta-\sqrt{\Delta^2-m^2+i\epsilon}} \right)  
%\eeq  
These are shown in Fig.~\ref{fig:kappav r1 r2} and the values obtained  
at the physical point are in agreement with experiment, which again indicates  
that the chiral extrapolation of the form factors could bring lattice  
%%%TOM: in -> into; brings -> could bring  
data into agreement with experiment.

Alternately, fitting only $\kappa^{p-n}$ using three parameters, namely  
$\kappa_{\rm v}(0)$, $c_V$ and $E_1(\lambda=0.6\,{\rm GeV})$ and fixing  
$c_A=1.125$ yields  $\chi^2/{\rm d.o.f.}=0.5$ and provides a nice fit to the   
results on $\kappa^{p-n}$. The Dirac radius $r_1^{2\,p-n}$ has only one  
fit parameter, whereas the combination $r_2^{2\, p-n}\kappa^{p-n}$ at   
leading one-loop order would be predicted since the term  
proportional to $B_{c2}$ would be absent.  
However, one can allow for such a term, which parametrizes the short-distance contributions  
%%%TOM: analogous -> analogously  
to the Pauli radius and which can be regarded analogously to $B_{10}(\lambda)$  
in the Dirac radius~\cite{Gockeler:2003ay}. We perform a fit to $r_1^{2\,p-n}$ with fit parameter $B_{10}$
and to $r_2^{2\, p-n}\kappa^{p-n}$ with fit parameter $B_{c2}$ using the expressions of Eq.~(\ref{r1 r2}).
The resulting values of the parameters from fitting  $\kappa^{p-n}$, $r_1^{2\,p-n}$ and  $r_2^{2\, p-n}\kappa^{p-n}$ independently
 are given in Table~\ref{tab:fit params}.  
%and displayed in Figs.~\ref{kappav r1 r2}.  
%%%TOM: combine fit -> combined fit  
If one treats $c_A$  as a fit parameter and performs a combined fit with  
six  parameters to  
$\kappa^{p-n}$, $r_1^{2\,p-n}$ and $r_2^{2\, p-n}\kappa^{p-n}$ one  
obtains   $\chi^2/{\rm d.o.f}=1.6$.   
The resulting fit for $\kappa^{p-n}$ is the same as that obtained by fitting  
separately the data on $\kappa^{p-n}$.   
The value of $\kappa^{p-n}$  at the physical point extracted from  
fitting the $F_1$ and $F_2$ agrees best with   
 the physical value. This is also true for the Dirac radius, whereas   
for the Pauli radius the deviation from the physical value is larger.  
Omitting the term proportional to $B_{c2}$ in the Pauli radius increases  
%%%TOM: expect -> except  
the value of  $\chi^2/{\rm d.o.f}$ and the resulting fits are similar except   
for the Dirac radius above pions of about 300~MeV.  
%If one does not include the term  
%proportional to $B_{c2}$ in the Pauli radius, the fit yields  
%a higher $chi^2/{\rm d.o.f.}=4.2$,  mainly due to the fact that  
%the radii are not reproduced, whereas  
%for $\kappa^{p-n}$ the fit is similar to the one obtained when  
%fitted separately.   

%%%TOM: at -> in  
\section{Results in the continuum limit}  
\begin{figure}  
\includegraphics[width=\linewidth]{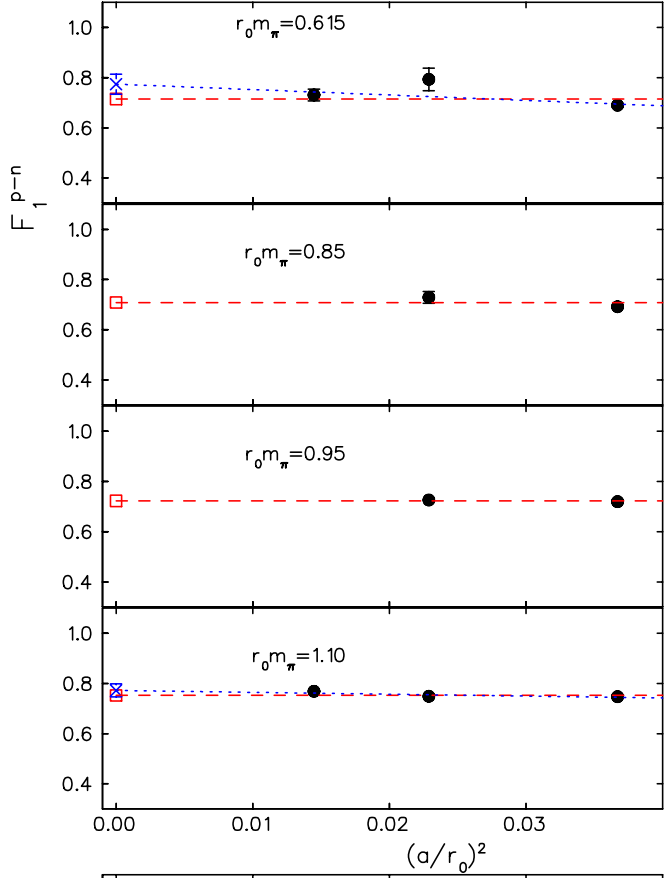}  
\includegraphics[width=\linewidth]{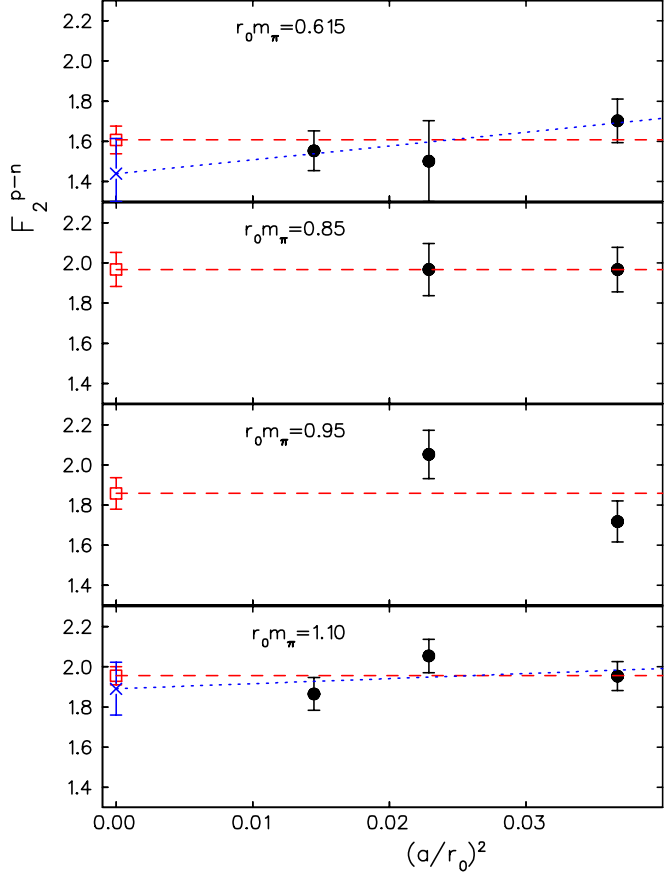}  
\caption{Continuum results for the isovector Dirac and Pauli  form factors  
$F_1$ and $F_2$ at $Q^2=357$ GeV$^2$.   
The dashed line is   
a fit to a constant, whereas the dotted line is a fit to a line.  
}  
\label{fig:continuum F1F2}  
\end{figure}

In order to study the dependence on the lattice  
spacing quantitatively we use the simulations at three lattice  
spacings at the smallest and largest pion mass used in this work. We take   
as reference pion mass the one computed on the finest lattice and  
interpolate results at the other two $\beta$-values to these two reference masses.  
In Fig.~\ref{fig:continuum F1F2} we show the value of the Dirac and Pauli  
$F_1$ and $F_2$ at these reference pion  
masses computed in units of $r_0$. We note that we first  
%%%TOM: at -> to  
interpolate these form factors to the same value of $Q^2$.   
In the figure we show the form factors at $Q^2=0.357$~GeV. We perform a fit to  
these data using   a linear form $F_1(a^2)=F1(0)+c(a/r_0)^2$. The resulting  
fit is shown in Fig.~\ref{fig:continuum F1F2}. Setting $c=0$ we obtain  
the constant line also shown in the figure. As can be seen, for both large  
and small pion masses the slope is consistent with zero yielding a value  
in the continuum limit in agreement with the constant fit.  
Therefore, we conclude that finite $a$ effects are negligible and  
for the intermediate pion masses we   
obtain the values in the continuum by fitting our data at $\beta=3.9$ and $\beta=4.05$ to a constant.  
    
  \begin{table}[h]  
\begin{tabular}{c|c|c|c|c}  
\hline\hline  
$r_0m_\pi$  & \multicolumn{4}{c} {$\kappa^{p-n}$ }  \\  
             & $(\beta=3.9)$   &   $(\beta=4.05)$  & $(\beta=4.2)$ & $(a\rightarrow 0)$ \\  
\hline   
 1.1019& 2.12(13) & 2.13(15) & 1.84(17) & 2.05(8) [1.79(26)]   \\  
 1.0   & 2.08(18) & 2.36(23) &          & 2.18(14)   \\  
 0.95  & 2.09(21) & 2.42(22) &          & 2.25(15)    \\  
 0.85  & 2.27(27) & 2.53(26) &          & 2.41(19)    \\  
 0.686 & 2.34(36) & 2.54(51) &          & 2.40(29)     \\  
 0.615 & 2.49(23) & 2.57(53) & 2.62(24) & 2.55(16) [2.71(41)]    \\\hline  
\end{tabular}  
\caption{In the second, third and fourth column we give the interpolated values of $\kappa^{p-n}$   
at the value of $m_\pi r_0$ given in the first column. We used $r_0/a=5.22(2)$, $6.61(3)$ and $8.31(5)$ for $\beta=3.9$, $4.05$ and $4.2$, respectively. In the fifth column we give the value of $\kappa^{p-n}$  after extrapolating to $a=0$ using a constant fit. In the parenthesis we give the corresponding values when using a linear fit. }  
\label{Table:kappa_continuum}  
\end{table}

\begin{figure}  
\includegraphics[width=\linewidth]{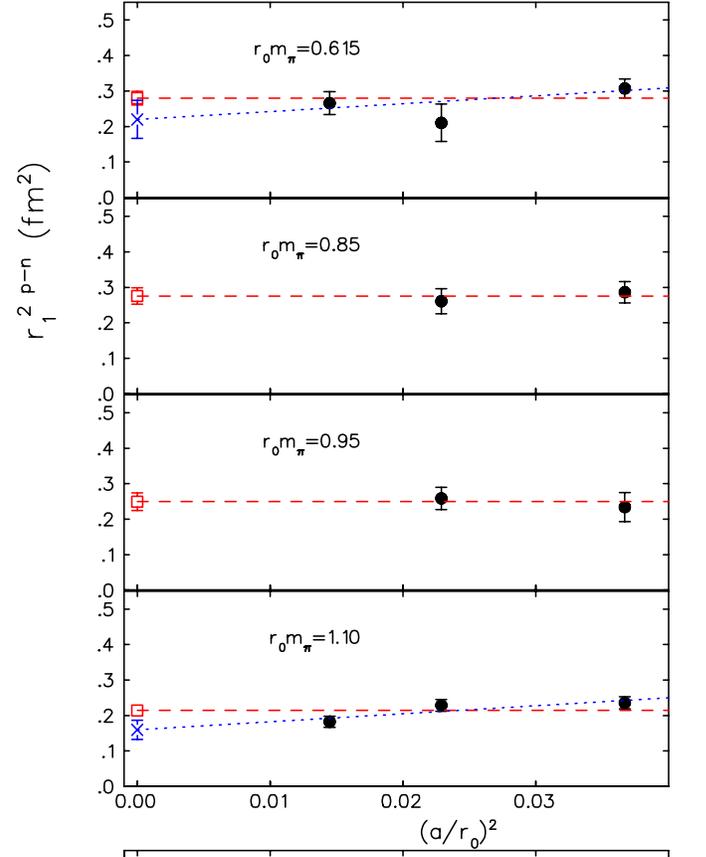}  
\includegraphics[width=\linewidth]{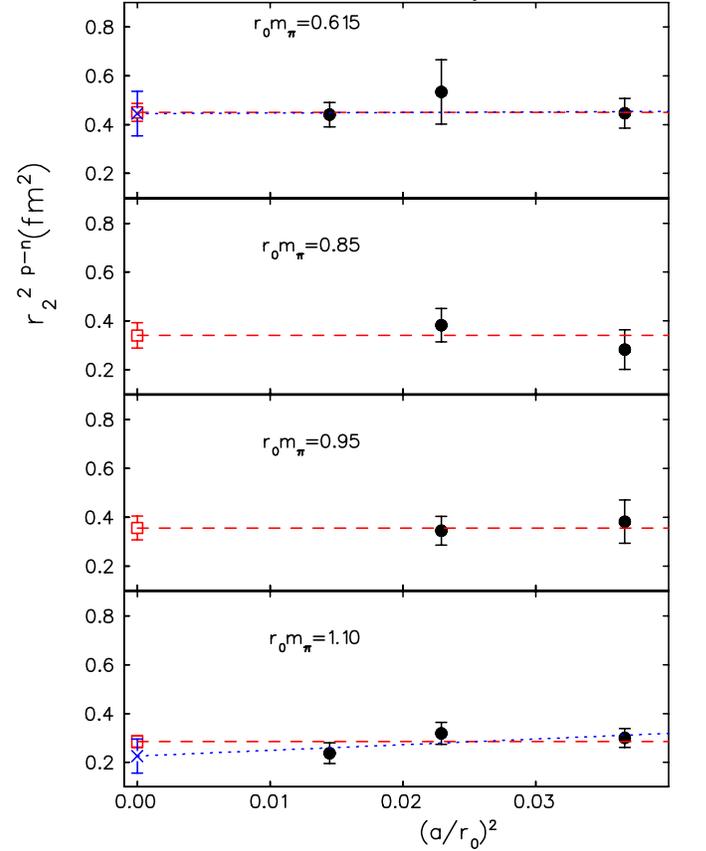}  
\caption{Continuum results for the isovector Dirac and Pauli mean squared radii $r_1^{2 \,p-n}$ and $r_2^{2\, p-n}$. The notation is the same as in Fig.~\ref{fig:continuum F1F2}.  
}  
\label{fig:r1r2 continuum}  
\end{figure}

\begin{figure}  
\includegraphics[width=\linewidth]{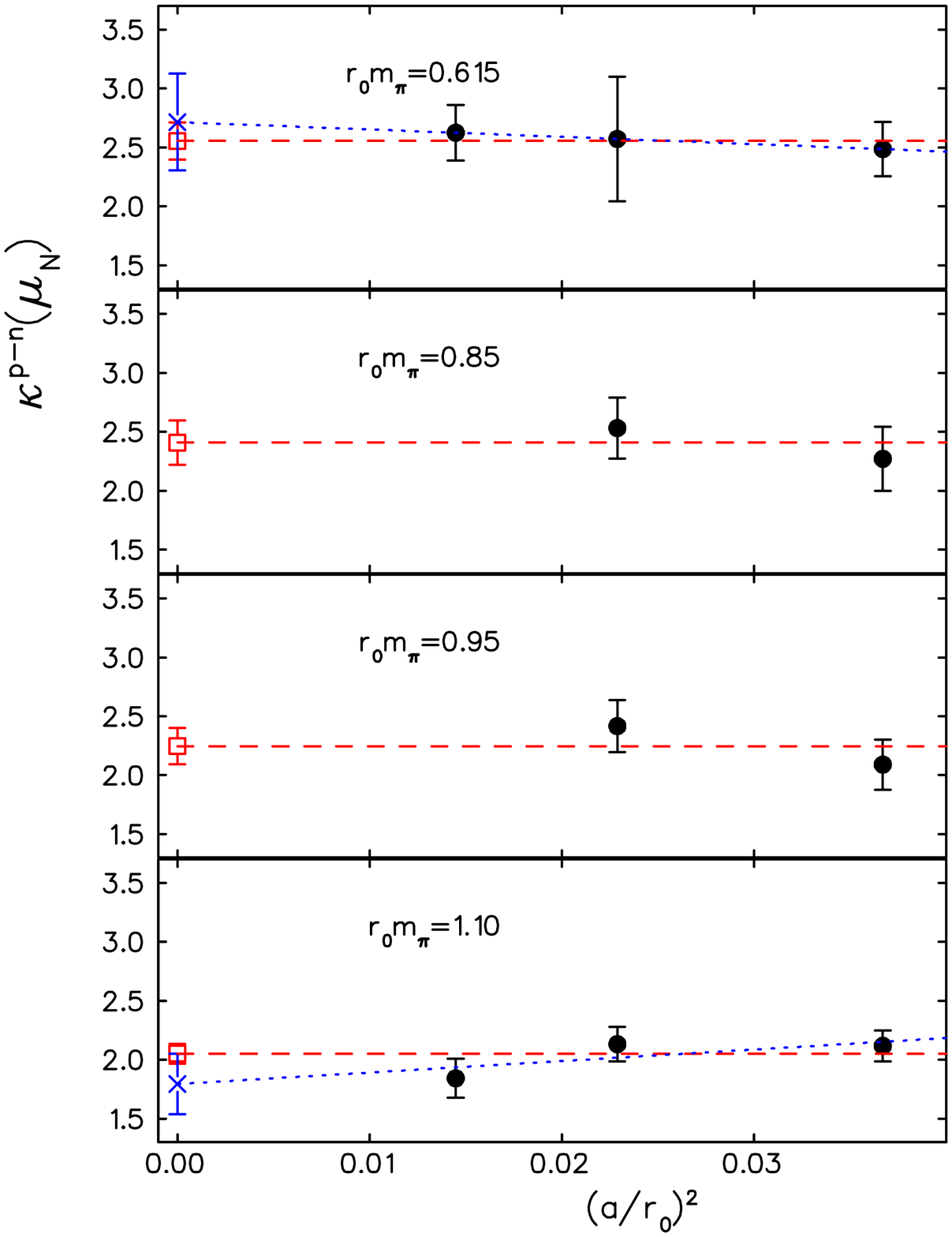}  
\caption{Continuum results for the nucleon  anomalous magnetic moment $\kappa^{p-n}$. The notation is the same as in Fig.~\ref{fig:continuum F1F2}.  
}  
\label{fig:kappav continuum}  
\end{figure}

In Figs.~\ref{fig:r1r2 continuum} and .~\ref{fig:kappav continuum} a we show the continuum extrapolation of   the r.m.s  Dirac and Pauli radii and the anomalous magnetic moment, respectively.    
The corresponding  values  of $\kappa^{p-n}$ at the six reference pion masses used in the figures are given in Table~\ref{Table:kappa_continuum} and those of the Dirac and Pauli mean square radii in Table~\ref{Table:r_continuum}.   
  
\begin{figure}[h] 
\includegraphics[width=\linewidth]{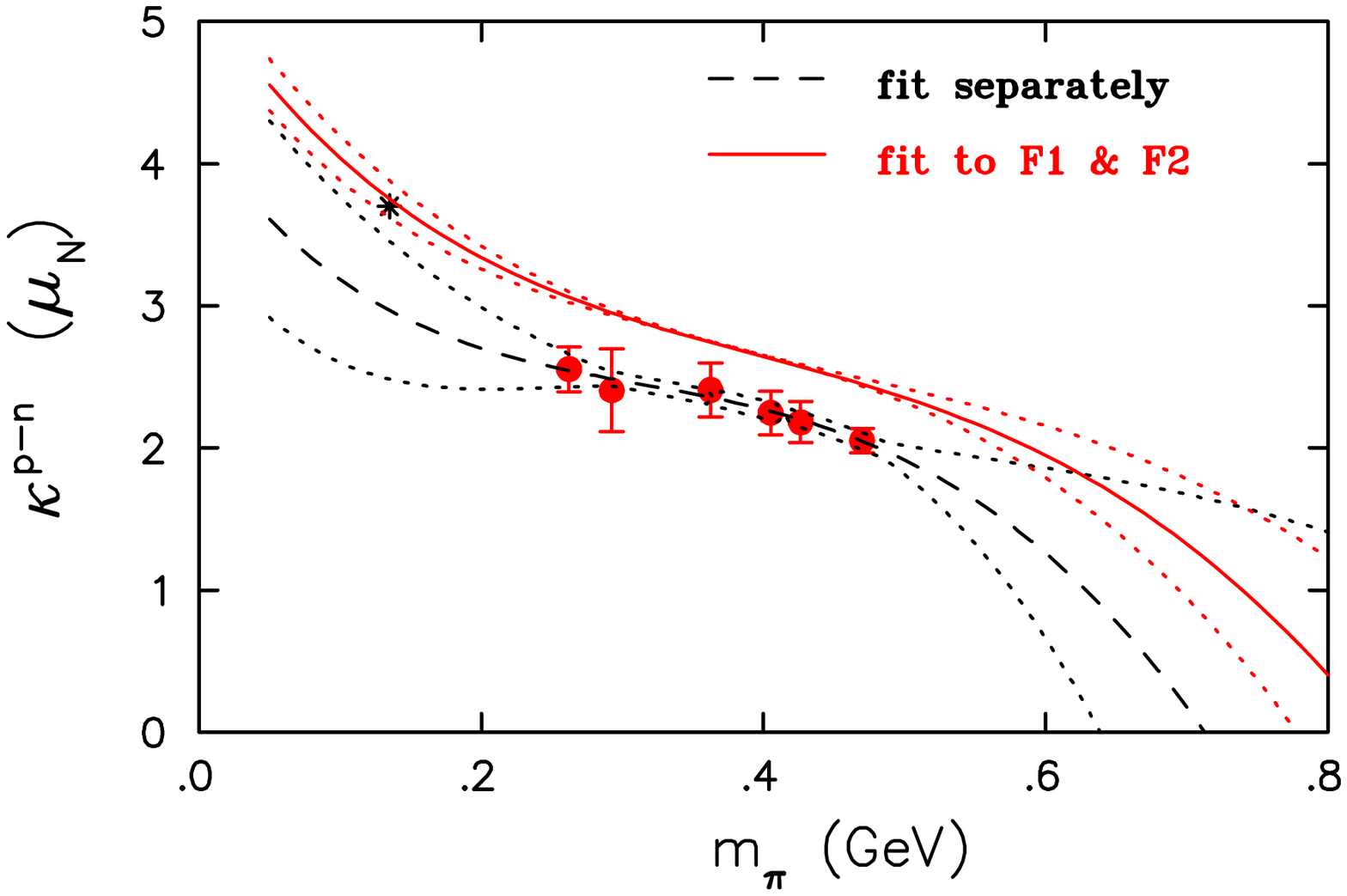}  
\caption{Chiral extrapolation of the results on the nucleon  anomalous magnetic moment $\kappa^{p-n}$ after taking the continuum limit. The dashed line  
is obtained by fitting $\kappa^{p-n}$ independently, whereas the solid  
line is the result of fitting the Dirac and Pauli form factors.}  
\label{fig:kappav cont}  
\end{figure}

\begin{figure}[h]  
\includegraphics[width=\linewidth]{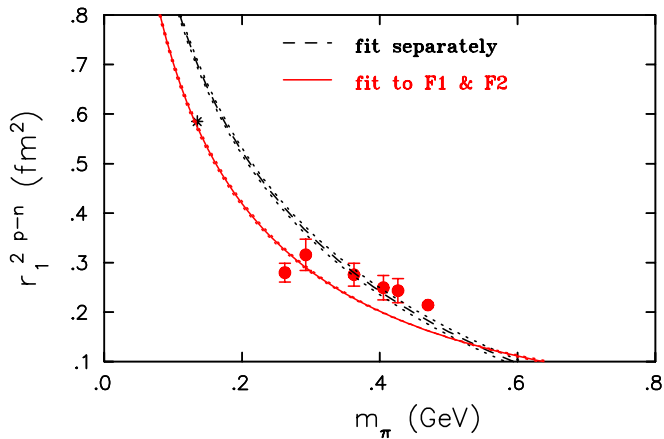}  
\caption{As in Fig.~\ref{fig:kappav cont} but for $r_1^{2\, p-n}$.}  
\label{fig:r1 cont}  
\end{figure}

\begin{figure}[h]\vspace*{0.5cm}  
\includegraphics[width=\linewidth]{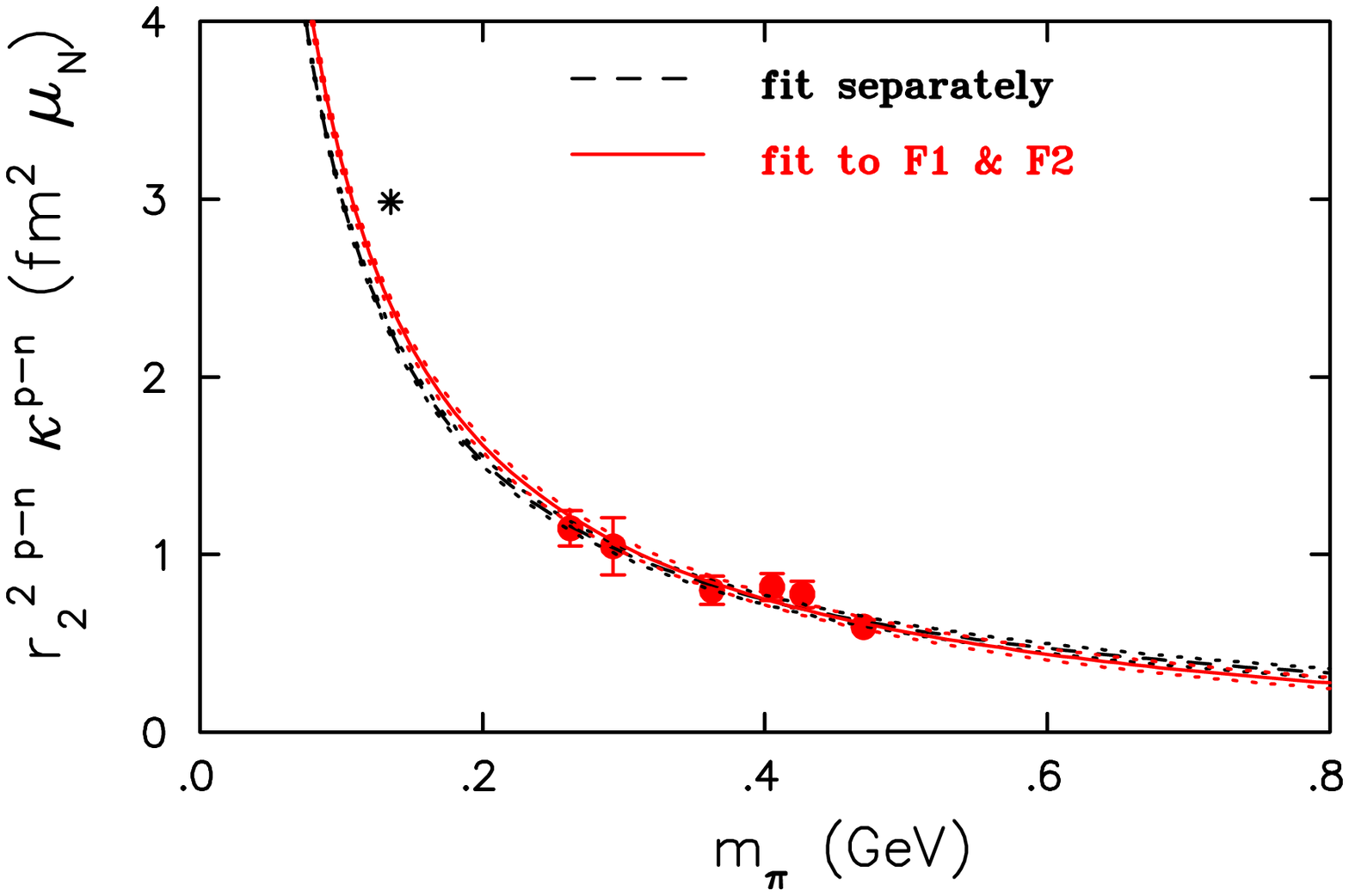}  
\caption{As in Fig.~\ref{fig:kappav cont} but for $r_2^{2\, p-n}\kappa^{p-n}$.}  
\label{fig:r2 cont}  
\end{figure}

%%%TOM: at->in  
Having results in the  
continuum limit we can now  
perform the chiral fits described in the previous section.  
We  show these chiral fits  
to the continuum results for  
the anomalous magnetic moment and Dirac and Pauli mean square radii  
 in Figs.~\ref{fig:kappav cont}, \ref{fig:r1 cont} and \ref{fig:r2 cont}.


The behavior observed is similar to that obtained when using the raw lattice  
data. Namely, chiral fits to the Dirac and Pauli form factors $F_1$ and $F_2$  
bring agreement with experiment at low $Q^2$-values, and therefore  
the  values for $\kappa^{p-n}$ and $r_1^{2\, p-n}$  derived  
using the parameters of the chiral fit to $F_1$ and $F_2$ agree with the  
  the experimental values.  The description of  $r_2^{2\, p-n}$ is also
reasonable bringing lattice
results close  to the value obtained at the physical point, although
not fully reproducing the experimental value.
%%%TOM: we are include -> we include; the the -> the; fitting the separately -> fitting separately  
In the figures we also include the curves obtained by fitting  
separately the anomalous magnetic moment and radii. For the former the  
mean value obtained at the physical point is lower as compared to the
 value obtained from fitting $F_1$ and $F_2$, with, however, almost overlapping
errors.   
For the Pauli mean squared radius the fits are almost identical, whereas for the   
%%%TOM: as compared to -> than  
Dirac radius both fits do not provide a good description to our lattice results.  
The parameters extracted from the chiral fits to the continuum extrapolated  
results are in fact in agreement  
with those determined using lattice data at the different $\beta$-values,  
as can be seen from the values given in Table~\ref{tab:fit params}.  
This is an {\it a posteriori} justification of using the continuum HB perturbation  
expressions to fit the lattice data at finite lattice spacing.  
\begin{widetext}  
\begin{center}  
\begin{table}[h]  
\begin{tabular}{c|cc|cc|cc|cc}  
\hline\hline  
$r_0m_\pi$  &$r_1^{2\,p-n}$ &  $r_2^{2\, p-n}$  &$r_1^{2\,p-n}$ &  $r_2^{2\, p-n}$ & $r_1^{2\,p-n}$ &  $r_2^{2\, p-n}$& $r_1^{2\,p-n}$ &  $r_2^{2\, p-n}$\\  
 & \multicolumn{2}{c}{$(\beta=3.9)$} &   \multicolumn{2}{c}{$(\beta=4.05)$} &  \multicolumn{2}{c}{$(\beta=4.2)$} &  \multicolumn{2}{c}{$(a\rightarrow 0)$}\\\hline   
 1.1019& 0.236(17)&0.300(39) &0.229(17) & 0.319(45)  &0.183(16)  & 0.238(43)& 0.214(9)[0.160(39)]  & 0.285(24) [0.226(70)]\\  
 1.0   & 0.226(36)&0.379(79) &0.258(33) & 0.326(62)  &           &          & 0.243(24)            & 0.347(49) \\  
 0.95  & 0.234(41)&0.382(89) &0.259(32) & 0.345(59)  &           &          & 0.249(25)            & 0.356(49)  \\  
 0.85  & 0.286(30)&0.282(81) &0.261(35) & 0.382(68)  &           &          & 0.276(23)            & 0.340(52)  \\  
 0.686 & 0.378(41)&0.409(99) &0.221(50) & 0.494(126) &           &          & 0.442(78)            & 0.442(78)\\  
 0.615 & 0.307(27)&0.446(61) &0.210(53) & 0.534(132) & 0.266(32) & 0.441(50)& 0.280(17) [0.220(54)]& 0.450(37) [0.445(91)]\\  
\hline  
\end{tabular}  
\caption{In the second, third and fourth raws we give the interpolated values of $r_1^{2\,p-n}$ and  
 $r_2^{2\,p-n}$in fm$^2$  
at the value of $m_\pi r_0$ given in the first column. We used $r_0/a=5.22(2)$, $6.61(3)$ and $8.31(5)$ for $\beta=3.9$, $4.05$ and $4.2$, respectively. In the fifth raws we give the value of $r_1^{2\, p-n}$   
and $r_2^{2\, p-n}$  after extrapolating to $a=0$ using a constant fit. In the parenthesis we give the corresponding values when using a linear fit. }  
\label{Table:r_continuum}  
\end{table}  
\end{center}  
\end{widetext}  
  
\section{Conclusions}  
Computing the electromagnetic form factors of the nucleon directly from the  
fundamental theory of the strong interactions has been the goal of hadron physics since the discovery of QCD.   
Within the lattice formulation this goal is now being realized.  
Comparing lattice results with a number of different   
fermion discretization schemes we find an overall agreement.   
In this work we   
 use dynamical simulations of two-degenerate flavors of light quarks  
in the twisted mass formulation of QCD, which  
at maximal twist is automatically ${\cal O}(a^2)$ improved, thus  
requiring no improvement on the operator level.  
We use light quark masses yielding pion masses in the range of   
%%%TOM: inserted 'our lightest'  
 about 260~MeV up to 470~MeV. Even for our lightest pion mass of 260~MeV the  
form factors decrease slower with increasing momentum transfer  
%%%TOM: as compared to -> than form factors obtained from experiments  
squared than form factors obtained from experiments. In this work  
we examine both volume and cut-off effects to identify the  
source of this discrepancy. We also examine the pion mass dependence  
of the form factors as well as of the quantities derived by fitting the  
$Q^2$-dependence of these form factors.  
  
By comparing results at two different volumes we find that for  
$Lm_\pi\stackrel{>}{\sim}3.3$ any volume effects are within our   
statistical accuracy for the magnetic form factor.
A small volume dependence is seen in the case of the electric form factor
that indicates an increase in the slope as the volume increases. 
 By considering the continuum   
limit using results at three lattice spacings we also show that,   
cut-off effects are small for lattice spacings less than about 0.1~fm.  
The pion mass dependence is examined using HB effective theory with   
explicit $\Delta$-degrees of freedom. Fitting the isovector Dirac and  
Pauli form factors at low $Q^2$ we show that the chiral extrapolated data   
agree with experiment. This is true when using the lattice results  
at the three $\beta$-values as well as when using the lattice  
results after taking the continuum limit.

\section*{Acknowledgments}  
We would like to thank all members of ETMC for a  
very constructive and enjoyable collaboration and for the many fruitful   
discussions that took place during the development of this work.  
  
Numerical calculations have  used HPC resources from GENCI Grant 2010-052271 (i.o. 2009)  and  
CC-IN2P3 as well as  from the  
John von Neumann-Institute for Computing on the JUMP and  Jugene systems at the   
research center  
in J\"ulich. We thank the staff members for their kind and sustained support.   
This work is supported in part by  the DFG  
Sonder\-for\-schungs\-be\-reich/ Trans\-regio SFB/TR9 and by funding received from the  
 Cyprus Research Promotion Foundation under contracts EPYAN/0506/08,  
KY-$\Gamma$/0907/11/ and TECHNOLOGY/$\Theta$E$\Pi$I$\Sigma$/0308(BE)/17.   
\bibliography{ETMC_ref}  

\begin{thebibliography}{32}
\expandafter\ifx\csname natexlab\endcsname\relax\def\natexlab#1{#1}\fi
\expandafter\ifx\csname bibnamefont\endcsname\relax
  \def\bibnamefont#1{#1}\fi
\expandafter\ifx\csname bibfnamefont\endcsname\relax
  \def\bibfnamefont#1{#1}\fi
\expandafter\ifx\csname citenamefont\endcsname\relax
  \def\citenamefont#1{#1}\fi
\expandafter\ifx\csname url\endcsname\relax
  \def\url#1{\texttt{#1}}\fi
\expandafter\ifx\csname urlprefix\endcsname\relax\def\urlprefix{URL }\fi
\providecommand{\bibinfo}[2]{#2}
\providecommand{\eprint}[2][]{\url{#2}}

\bibitem[{\citenamefont{Jones et~al.}(2000)}]{Jones:1999rz}
\bibinfo{author}{\bibfnamefont{M.~K.} \bibnamefont{Jones}} \bibnamefont{et~al.}
  (\bibinfo{collaboration}{Jefferson Lab Hall A}), \bibinfo{journal}{Phys. Rev.
  Lett.} \textbf{\bibinfo{volume}{84}}, \bibinfo{pages}{1398}
  (\bibinfo{year}{2000}), \eprint{nucl-ex/9910005}.

\bibitem[{\citenamefont{Gayou et~al.}(2002)}]{Gayou:2001qd}
\bibinfo{author}{\bibfnamefont{O.}~\bibnamefont{Gayou}} \bibnamefont{et~al.}
  (\bibinfo{collaboration}{Jefferson Lab Hall A}), \bibinfo{journal}{Phys. Rev.
  Lett.} \textbf{\bibinfo{volume}{88}}, \bibinfo{pages}{092301}
  (\bibinfo{year}{2002}), \eprint{nucl-ex/0111010}.

\bibitem[{\citenamefont{Perdrisat et~al.}(2007)\citenamefont{Perdrisat,
  Punjabi, and Vanderhaeghen}}]{Perdrisat:2006hj}
\bibinfo{author}{\bibfnamefont{C.~F.} \bibnamefont{Perdrisat}},
  \bibinfo{author}{\bibfnamefont{V.}~\bibnamefont{Punjabi}}, \bibnamefont{and}
  \bibinfo{author}{\bibfnamefont{M.}~\bibnamefont{Vanderhaeghen}},
  \bibinfo{journal}{Prog. Part. Nucl. Phys.} \textbf{\bibinfo{volume}{59}},
  \bibinfo{pages}{694} (\bibinfo{year}{2007}), \eprint{hep-ph/0612014}.

\bibitem[{\citenamefont{Guichon and Vanderhaeghen}(2003)}]{Guichon:2003qm}
\bibinfo{author}{\bibfnamefont{P.~A.~M.} \bibnamefont{Guichon}}
  \bibnamefont{and}
  \bibinfo{author}{\bibfnamefont{M.}~\bibnamefont{Vanderhaeghen}},
  \bibinfo{journal}{Phys. Rev. Lett.} \textbf{\bibinfo{volume}{91}},
  \bibinfo{pages}{142303} (\bibinfo{year}{2003}), \eprint{hep-ph/0306007}.

\bibitem[{\citenamefont{de~Jager}(2008)}]{deJager:2008zza}
\bibinfo{author}{\bibfnamefont{K.}~\bibnamefont{de~Jager}},
  \bibinfo{journal}{Nucl. Phys.} \textbf{\bibinfo{volume}{A805}},
  \bibinfo{pages}{494} (\bibinfo{year}{2008}).

\bibitem[{\citenamefont{Shindler}(2008)}]{Shindler:2007vp}
\bibinfo{author}{\bibfnamefont{A.}~\bibnamefont{Shindler}},
  \bibinfo{journal}{Phys. Rept.} \textbf{\bibinfo{volume}{461}},
  \bibinfo{pages}{37} (\bibinfo{year}{2008}), \eprint{0707.4093}.

\bibitem[{\citenamefont{Frezzotti et~al.}(2001)\citenamefont{Frezzotti, Grassi,
  Sint, and Weisz}}]{Frezzotti:2000nk}
\bibinfo{author}{\bibfnamefont{R.}~\bibnamefont{Frezzotti}},
  \bibinfo{author}{\bibfnamefont{P.~A.} \bibnamefont{Grassi}},
  \bibinfo{author}{\bibfnamefont{S.}~\bibnamefont{Sint}}, \bibnamefont{and}
  \bibinfo{author}{\bibfnamefont{P.}~\bibnamefont{Weisz}}
  (\bibinfo{collaboration}{Alpha}), \bibinfo{journal}{JHEP}
  \textbf{\bibinfo{volume}{0108}}, \bibinfo{pages}{058} (\bibinfo{year}{2001}),
  \eprint{hep-lat/0101001}.

\bibitem[{\citenamefont{Frezzotti and Rossi}(2004)}]{Frezzotti:2003ni}
\bibinfo{author}{\bibfnamefont{R.}~\bibnamefont{Frezzotti}} \bibnamefont{and}
  \bibinfo{author}{\bibfnamefont{G.~C.} \bibnamefont{Rossi}},
  \bibinfo{journal}{JHEP} \textbf{\bibinfo{volume}{08}}, \bibinfo{pages}{007}
  (\bibinfo{year}{2004}), \eprint{hep-lat/0306014}.

\bibitem[{\citenamefont{Weisz}(1983)}]{Weisz:1982zw}
\bibinfo{author}{\bibfnamefont{P.}~\bibnamefont{Weisz}},
  \bibinfo{journal}{Nucl. Phys.} \textbf{\bibinfo{volume}{B212}},
  \bibinfo{pages}{1} (\bibinfo{year}{1983}).

\bibitem[{\citenamefont{Alexandrou}(2009)}]{Alexandrou:2009xk}
\bibinfo{author}{\bibfnamefont{C.}~\bibnamefont{Alexandrou}}
  (\bibinfo{year}{2009}), \eprint{0906.4137}.

\bibitem[{\citenamefont{Alexandrou
  et~al.}(2009{\natexlab{a}})}]{Alexandrou:2009qu}
\bibinfo{author}{\bibfnamefont{C.}~\bibnamefont{Alexandrou}}
  \bibnamefont{et~al.} (\bibinfo{collaboration}{ETM}), \bibinfo{journal}{Phys.
  Rev.} \textbf{\bibinfo{volume}{D80}}, \bibinfo{pages}{114503}
  (\bibinfo{year}{2009}{\natexlab{a}}), \eprint{0910.2419}.

\bibitem[{\citenamefont{Drach et~al.}(2008)}]{Drach:2009dh}
\bibinfo{author}{\bibfnamefont{V.}~\bibnamefont{Drach}} \bibnamefont{et~al.},
  \bibinfo{journal}{PoS} \textbf{\bibinfo{volume}{LATTICE2008}},
  \bibinfo{pages}{123} (\bibinfo{year}{2008}), \eprint{0905.2894}.

\bibitem[{\citenamefont{Alexandrou
  et~al.}(2008{\natexlab{a}})}]{Alexandrou:2008tn}
\bibinfo{author}{\bibfnamefont{C.}~\bibnamefont{Alexandrou}}
  \bibnamefont{et~al.} (\bibinfo{collaboration}{European Twisted Mass}),
  \bibinfo{journal}{Phys. Rev.} \textbf{\bibinfo{volume}{D78}},
  \bibinfo{pages}{014509} (\bibinfo{year}{2008}{\natexlab{a}}),
  \eprint{0803.3190}.

\bibitem[{\citenamefont{Alexandrou et~al.}(2007)}]{Alexandrou:2007qq}
\bibinfo{author}{\bibfnamefont{C.}~\bibnamefont{Alexandrou}}
  \bibnamefont{et~al.} (\bibinfo{collaboration}{ETM Collaboration}),
  \bibinfo{journal}{PoS} \textbf{\bibinfo{volume}{LAT2007}},
  \bibinfo{pages}{087} (\bibinfo{year}{2007}), \eprint{arXiv:0710.1173
  [hep-lat]}.

\bibitem[{\citenamefont{Drach et~al.}(2010)}]{Drach:2010}
\bibinfo{author}{\bibfnamefont{V.}~\bibnamefont{Drach}} \bibnamefont{et~al.},
  \bibinfo{journal}{PoS} \textbf{\bibinfo{volume}{Lattice 2010}},
  \bibinfo{pages}{123} (\bibinfo{year}{2010}).

\bibitem[{\citenamefont{Alexandrou}(2010)}]{Alexandrou:2010}
\bibinfo{author}{\bibfnamefont{C.}~\bibnamefont{Alexandrou}}
  (\bibinfo{collaboration}{ETM Collaboration}), \bibinfo{journal}{PoS}
  \textbf{\bibinfo{volume}{Lattice 2010}}, \bibinfo{pages}{001}
  (\bibinfo{year}{2010}).

\bibitem[{\citenamefont{Alexandrou
  et~al.}(2009{\natexlab{b}})}]{Alexandrou:2009ng}
\bibinfo{author}{\bibfnamefont{C.}~\bibnamefont{Alexandrou}}
  \bibnamefont{et~al.}, \bibinfo{journal}{PoS}
  \textbf{\bibinfo{volume}{LAT2009}}, \bibinfo{pages}{145}
  (\bibinfo{year}{2009}{\natexlab{b}}), \eprint{0910.3309}.

\bibitem[{\citenamefont{Alexandrou
  et~al.}(2008{\natexlab{b}})}]{Alexandrou:2008rp}
\bibinfo{author}{\bibfnamefont{C.}~\bibnamefont{Alexandrou}}
  \bibnamefont{et~al.}, \bibinfo{journal}{PoS LAT2008}
  \textbf{\bibinfo{volume}{B414}}, \bibinfo{pages}{145}
  (\bibinfo{year}{2008}{\natexlab{b}}), \eprint{hep-lat/9211042}.

\bibitem[{\citenamefont{Alexandrou et~al.}(1994)\citenamefont{Alexandrou,
  Gusken, Jegerlehner, Schilling, and Sommer}}]{Alexandrou:1992ti}
\bibinfo{author}{\bibfnamefont{C.}~\bibnamefont{Alexandrou}},
  \bibinfo{author}{\bibfnamefont{S.}~\bibnamefont{Gusken}},
  \bibinfo{author}{\bibfnamefont{F.}~\bibnamefont{Jegerlehner}},
  \bibinfo{author}{\bibfnamefont{K.}~\bibnamefont{Schilling}},
  \bibnamefont{and} \bibinfo{author}{\bibfnamefont{R.}~\bibnamefont{Sommer}},
  \bibinfo{journal}{Nucl. Phys.} \textbf{\bibinfo{volume}{B414}},
  \bibinfo{pages}{815} (\bibinfo{year}{1994}), \eprint{hep-lat/9211042}.

\bibitem[{\citenamefont{Gusken}(1990)}]{Gusken:1989}
\bibinfo{author}{\bibfnamefont{S.}~\bibnamefont{Gusken}},
  \bibinfo{journal}{Nucl. Phys. Proc. Suppl.} \textbf{\bibinfo{volume}{17}},
  \bibinfo{pages}{361} (\bibinfo{year}{1990}).

\bibitem[{\citenamefont{Alexandrou et~al.}(2006)\citenamefont{Alexandrou,
  Koutsou, Negele, and Tsapalis}}]{Alexandrou:2006ru}
\bibinfo{author}{\bibfnamefont{C.}~\bibnamefont{Alexandrou}},
  \bibinfo{author}{\bibfnamefont{G.}~\bibnamefont{Koutsou}},
  \bibinfo{author}{\bibfnamefont{J.~W.} \bibnamefont{Negele}},
  \bibnamefont{and} \bibinfo{author}{\bibfnamefont{A.}~\bibnamefont{Tsapalis}},
  \bibinfo{journal}{Phys. Rev.} \textbf{\bibinfo{volume}{D74}},
  \bibinfo{pages}{034508} (\bibinfo{year}{2006}), \eprint{hep-lat/0605017}.

\bibitem[{\citenamefont{Alexandrou et~al.}(2010)}]{Alexandrou:2010hf}
\bibinfo{author}{\bibfnamefont{C.}~\bibnamefont{Alexandrou}}
  \bibnamefont{et~al.} (\bibinfo{collaboration}{ETM}) (\bibinfo{year}{2010}),
  \eprint{1012.0857}.

\bibitem[{\citenamefont{Urbach}(2007)}]{Urbach:2007}
\bibinfo{author}{\bibfnamefont{C.}~\bibnamefont{Urbach}},
  \bibinfo{journal}{PoS} \textbf{\bibinfo{volume}{LAT2007}},
  \bibinfo{pages}{022} (\bibinfo{year}{2007}).

\bibitem[{\citenamefont{Baron et~al.}(2010)}]{Baron:2009wt}
\bibinfo{author}{\bibfnamefont{R.}~\bibnamefont{Baron}} \bibnamefont{et~al.}
  (\bibinfo{collaboration}{ETM}), \bibinfo{journal}{JHEP}
  \textbf{\bibinfo{volume}{08}}, \bibinfo{pages}{097} (\bibinfo{year}{2010}),
  \eprint{0911.5061}.

\bibitem[{\citenamefont{Kelly}(2004)}]{Kelly:2004hm}
\bibinfo{author}{\bibfnamefont{J.~J.} \bibnamefont{Kelly}},
  \bibinfo{journal}{Phys. Rev.} \textbf{\bibinfo{volume}{C70}},
  \bibinfo{pages}{068202} (\bibinfo{year}{2004}).

\bibitem[{\citenamefont{Syritsyn et~al.}(2010)}]{Syritsyn:2009mx}
\bibinfo{author}{\bibfnamefont{S.~N.} \bibnamefont{Syritsyn}}
  \bibnamefont{et~al.}, \bibinfo{journal}{Phys. Rev.}
  \textbf{\bibinfo{volume}{D81}}, \bibinfo{pages}{034507}
  (\bibinfo{year}{2010}), \eprint{0907.4194}.

\bibitem[{\citenamefont{Capitani et~al.}(2010)\citenamefont{Capitani,
  Della~Morte, Knippschild, and Wittig}}]{Capitani:2010sg}
\bibinfo{author}{\bibfnamefont{S.}~\bibnamefont{Capitani}},
  \bibinfo{author}{\bibfnamefont{M.}~\bibnamefont{Della~Morte}},
  \bibinfo{author}{\bibfnamefont{B.}~\bibnamefont{Knippschild}},
  \bibnamefont{and} \bibinfo{author}{\bibfnamefont{H.}~\bibnamefont{Wittig}}
  (\bibinfo{year}{2010}), \eprint{1011.1358}.

\bibitem[{\citenamefont{Bratt et~al.}(2010)}]{Bratt:2010jn}
\bibinfo{author}{\bibfnamefont{J.~D.} \bibnamefont{Bratt}} \bibnamefont{et~al.}
  (\bibinfo{collaboration}{LHPC}), \bibinfo{journal}{Phys. Rev.}
  \textbf{\bibinfo{volume}{D82}}, \bibinfo{pages}{094502}
  (\bibinfo{year}{2010}), \eprint{1001.3620}.

\bibitem[{\citenamefont{Hemmert and Weise}(2002)}]{Hemmert:2002uh}
\bibinfo{author}{\bibfnamefont{T.~R.} \bibnamefont{Hemmert}} \bibnamefont{and}
  \bibinfo{author}{\bibfnamefont{W.}~\bibnamefont{Weise}},
  \bibinfo{journal}{Eur. Phys. J.} \textbf{\bibinfo{volume}{A15}},
  \bibinfo{pages}{487} (\bibinfo{year}{2002}), \eprint{hep-lat/0204005}.

\bibitem[{\citenamefont{Gockeler et~al.}(2005)}]{Gockeler:2003ay}
\bibinfo{author}{\bibfnamefont{M.}~\bibnamefont{Gockeler}} \bibnamefont{et~al.}
  (\bibinfo{collaboration}{QCDSF}), \bibinfo{journal}{Phys. Rev.}
  \textbf{\bibinfo{volume}{D71}}, \bibinfo{pages}{034508}
  (\bibinfo{year}{2005}), \eprint{hep-lat/0303019}.

\bibitem[{\citenamefont{Yamazaki et~al.}(2009)}]{Yamazaki:2009zq}
\bibinfo{author}{\bibfnamefont{T.}~\bibnamefont{Yamazaki}}
  \bibnamefont{et~al.}, \bibinfo{journal}{Phys. Rev.}
  \textbf{\bibinfo{volume}{D79}}, \bibinfo{pages}{114505}
  (\bibinfo{year}{2009}), \eprint{0904.2039}.

\bibitem[{\citenamefont{Collins et~al.}(2011)}]{Collins:2011gw}
\bibinfo{author}{\bibfnamefont{S.}~\bibnamefont{Collins}} \bibnamefont{et~al.}
  (\bibinfo{year}{2011}), \eprint{1101.2326}.

\end{thebibliography}
\end{document}